\documentclass[a4paper,11pt]{article}
\usepackage{graphicx}
\usepackage[space]{grffile}
\usepackage{latexsym}
\usepackage{textcomp}
\usepackage{longtable}
\usepackage{tabulary}
\usepackage{booktabs,array,multirow}
\usepackage{amsfonts,amsmath,amssymb}
\usepackage{natbib}
\usepackage{float}
\usepackage{url}
\usepackage{hyperref}
\hypersetup{colorlinks,linkcolor={blue},citecolor={red},urlcolor={red}}  
\usepackage{etoolbox}
\makeatletter
\usepackage{authblk}
\usepackage[utf8]{inputenc}
\usepackage[english]{babel}
\usepackage[top=1in, bottom=1in, left=1in, right=1in]{geometry}
\newcommand{\supplementarysection}{%
  \setcounter{figure}{0}% Reset figure counter
  \let\oldthefigure\thefigure% Capture figure numbering scheme
  \renewcommand{\thefigure}{S\oldthefigure}% Prefix figure number with S
  \section{Supplementary section}% Set supplementary section
  }
\title{Role of Bay of Bengal Low Pressure Systems in the Formation of Mid-Tropospheric Cyclones over the Arabian Sea and Western India}
\date{}
\author[1,2]{Pradeep Kushwaha \thanks{Corresponding author: Pradeep Kushwaha, pkushwaha@gmail.com}}
\author[1,2]{Jai Sukhatme}
\author[1,2]{Ravi S. Nanjundiah}
\affil[1]{Centre for Atmospheric and Oceanic Sciences, Indian Institute of Science, Bangalore, 560012, India}
\affil[2]{Divecha Centre for Climate Change, Indian Institute of Science, Bangalore, 560012, India}
\begin{document}
%\linenumbers
\maketitle
%\linenumbers
%\clearpage

\begin{abstract}
Arabian Sea Mid-Tropospheric Cyclones (MTCs), which are responsible for extreme rainfall events in western India, often coincide with monsoon low-pressure systems (LPS) over the Bay of Bengal. However, the role of Bay of Bengal LPSs in the formation of Arabian Sea MTCs remains unclear. This study utilizes the Weather Research and Forecasting Model (WRF) to investigate the atmospheric connection between the two basins.
By introducing a balanced bogus vortex over the Bay of Bengal, cyclonic systems are induced over the Arabian Sea in the majority of ensemble members, exhibiting characteristics consistent with observations. As the Bay of Bengal vortex moves westward, the middle tropospheric trough deepens, horizontal wind shear increases, the low-level Arabian Sea stable inversion layer weakens, and the middle troposphere moisture content increases over western India and the northeast Arabian Sea. Subsequently, MTC genesis occurs along the western edge of the trough within 2-4 days of model integration over the northeast Arabian Sea. Vorticity budget analysis highlights the critical role of vorticity advection and tilting during the initial 24 hours of MTC genesis, while vortex stretching becomes the dominant vorticity source during rapid intensification. To further substantiate these findings, a mechanism denial experiment is conducted using a real-world instance of coexistent Arabian Sea MTC and Bay of Bengal LPS, which was replicated in the model. In the experiment, conditions unfavorable for LPS genesis are created by cooling and drying the Bay of Bengal. The results demonstrate that when the Bay of Bengal LPS does not develop or intensify, the Arabian Sea MTC fails to form. This study presents compelling evidence for the significant influence of Bay of Bengal low-pressure systems on the formation of severe weather-inducing MTCs over the Arabian Sea and western India.
\end{abstract}

\section{Introduction}
% \justify
\noindent
Many extreme rainfall events over western India are associated with Mid-Tropospheric Cyclones (MTCs) \cite{carr1977mid,francis2006intense,choudhury2018phenomenological,ksn}. These MTCs are characterized by their maximum intensity in the middle troposphere and a relatively weak signature in the lower troposphere. Regions in western India, particularly Maharashtra and Gujarat, including densely populated cities like Mumbai, experience exceptionally heavy rain exceeding 500 mm during each monsoon season due to the influence of MTCs \cite{mapes2011heaviest,choudhury2018phenomenological,kushwaha2023classification}. Despite their significant contribution to the seasonal rainfall in this area and their crucial role in extreme rainfall events, the precise mechanisms behind the genesis of MTCs over the Arabian Sea and western India remain poorly understood.

\noindent Some preliminary insights into the prevailing conditions during MTC genesis over the Arabian Sea have been obtained from detailed observations of the July 1963 MTC during the International Indian Ocean Expedition. The U.S. Weather Bureau Research Flight Facility (RFF) aircraft, in collaboration with the Woods Hole Oceanographic Institution, extensively explored the Arabian Sea from the surface to about 14 km between June and July 1963, capturing the development of an MTC within this observational network \cite{miller1968iioe}. Specifically, after the establishment of the monsoon circulation, a warm-core monsoon low formed on June 24 around 15$^{\circ}$N, 97$^{\circ}$E in the Bay of Bengal. Over the next two days, this system moved northwestwards, crossed the coast of the East Indian state of Odisha on June 26, and remained stationary for the following three days over the east coast of India. During this time, Rawin reports from Mumbai and Ahmedabad showed an enhancement of cyclonic shear between the 700-500 hPa layers. The strengthening of westerlies south of 20$^{\circ}$N and the extension of easterlies down to 850 hPa north of 20$^{\circ}$N were also observed. Subsequently, the genesis of MTC was observed within the shear zone on June 28 over the Konkan coast and the northeast Arabian Sea.The formation of the MTC led to the development of an east-west trough in the middle troposphere, extending from the Arabian Sea to the Bay of Bengal through peninsular India, which coincided with a significant active phase of the Indian monsoon from July 2 to 10.

\noindent During the formation of the MTC, it was noted that desert air from the north and west was overriding the moist surface layer \cite{miller1968iioe}. Ship and aircraft observations revealed the presence of extensive stratocumulus over the Arabian Sea west of 65$^{\circ}$E, indicating the trapping of moisture in the lowest 2 km under a strong temperature inversion caused by the differential temperature advection of desert air over cold marine air \cite{narayanan2004role,das2007simulation,das2021characteristics}. A dropsonde released at 500 hPa during the RFF flight confirmed the presence of an inversion at 900 hPa. This low-level inversion over the northeast Arabian Sea disappeared as soon as the MTC intensified. Furthermore, a significant increase in moisture content within the 700 hPa to 500 hPa layer was observed during the genesis of the MTC \cite{miller1968iioe}.

\noindent Another crucial aspect noted during the genesis of the July 1963 MTC was the preexistence of a low-pressure system (LPS) over the Bay of Bengal and East India \cite{miller1968iioe}. The intensification and westward motion of the Bay of Bengal LPS were accompanied by the formation of a shear zone and trough over the Arabian Sea, which extended up to the Bay of Bengal. This precedence and coexistence of the Bay of Bengal LPS during the MTC gensis were further confirmed in three additional cases of MTCs \cite{miller1968iioe} --- specifically, a time series of vorticity over the Arabian Sea, central India, and the Bay of Bengal showed that vorticity and rainfall over the Bay of Bengal and Central India preceded rainfall over western India. Recent studies have analyzed heavy precipitating MTCs of Indian Meteorological Department monsoon summaries and have found that 90\% of them formed in the presence of a Bay of Bengal LPS \cite{choudhury2018phenomenological}. Additionally, detailed objective tracking and classification of MTCs over the Indian region revealed that 83\% of in-situ Arabian Sea MTC formation occurred with a preexisting LPS over Bay of Bengal \cite{kushwaha2023classification}. Moreover, the Arabian Sea MTCs that coexisted with the Bay of Bengal cyclonic systems constituted the rainiest class (Type-2a) of synoptic systems over western India \cite{kushwaha2022classification}.

\noindent While efforts have been made to understand MTC genesis in the Arabian Sea from the barotropic and baroclinic instability of the mean flow \cite{mak1975monsoonal,mak1982instability,goswami1980role}, aforementioned observational work has highlighted the significant occurrence of rainy MTCs in conjunction with preceding Bay of Bengal LPS \cite{carr1977mid,choudhury2018phenomenological,kushwaha2022classification}. However, it remains unclear whether the preexistence of cyclonic systems in the Bay of Bengal during the genesis of MTCs is merely coincidental or if there exists a dynamic linkage between the formation of systems in both basins. To address this, our study investigates the dynamical connection between MTC formation in the Arabian Sea and LPS or cyclonic  systems in the Bay of Bengal using reanalysis data and numerical experiments with the Weather Research and Forecasting Model (WRF). Gaining a comprehensive understanding of the role played by Bay of Bengal LPS in the formation of Arabian sea MTCs, as well as their impact on rainfall intensity and system maintenance, is of paramount importance for enhancing prediction capabilities and deepening our understanding of extreme rainfall events associated with MTCs. Such insights hold significant potential for advancing scientific knowledge and improving the accuracy of forecasts and warnings related to these meteorological phenomena. The outline of this paper is as follows: Section 2 presents the data used, while Section 3 describes the methods and model configuration. Section 4 presents the observational conditions and evolution of meteorological fields during the genesis of an MTC over the Arabian Sea and western India, occurring with a preceding Bay of Bengal LPS. In Section 5, we discuss numerical simulations that validate the role of Bay of Bengal LPSs in the formation of Arabian Sea MTCs. Finally, Section 6 concludes our study.
%================================================================
                    % DATA SECTION BEGIN
%================================================================
\section{Data}
\subsection{NCEP Final Analysis (FNL) Data}
WRF model is configured to ingest various atmosphere data sets to produce the initial and boundary conditions. We utilize the widely used National Centers for Environmental Prediction Final Analysis (NCEP-FNL) Global Analysis data set. FNL data are generated from the Global Data Assimilation System (GDAS), which uses observational data from the Global Telecommunications System (GTS) and other sources for assimilation. We utilized data at six hourly intervals and one-degree horizontal resolution, available at \url{https://rda.ucar.edu/datasets/ds083.2/}. In fact, this is one of the most commonly used data sets for research and forecasting purposes \citep{kalnay1996ncep,lo2008assessment,kesarkar2007coupling,zhang2011improved,melhauser2012practical}.

\subsection{ERA-5 Data}
\noindent The main reanalysis product used in this work as a proxy for observations is the ECMWF ERA-5 fifth-generation atmospheric reanalysis data set \cite{hersbach2020era5} which is generated using 41r2 of the Integrated Forecast System (IFS) model. IFS system utilizes a four-dimensional variational data assimilation scheme and takes advantage of 137 vertical levels and a horizontal resolution of $0.28125^{\circ}$ ($~31$ km, or TL639 triangular truncation). The data is stored at every hour of model integration. This study utilizes six-hourly winds, vorticity, divergence, temperature, and moisture fields on pressure levels from $1000-100$ hPa and native horizontal resolution. 
%However, we use interpolated data on 1.5 degrees latitude-longitude gird for synoptic charts.
Apart from a high spatial and temporal resolution, ERA-5 has several important updates to its predecessor ERA-I, which was terminated in 2019. These include ozone with satellite radiances, aircraft, and surface pressure data during the data assimilation. One of the critical changes in ERA-5 is using an all-sky approach instead of the clear-sky approach used in ERA-I, thus providing additional information about precipitation and cloud distribution. These updates and others have resulted in more consistent sea-surface temperature, and ice-sea compared to ERA-I \cite{hersbach2020era5}.

\subsection{Track data set}
For constructing composites of instances of coexisting Arabian Sea MTCs and the Bay of Bengal LPS, we use 60 Type-2a MTCs of \citet{ksn}. Type-2a are MTCs which form in the presence of a preceding Bay of Bengal LPS. %All observational composites will be shown for Type-2a systems.

\section{Methodology}

\subsection{Bogus Vortex Scheme}

\noindent To understand the effect of the Bay of Bengal LPS in the formation of Arabian Sea MTC, we superimposed a cyclonic vortex over the Bay of Bengal as a perturbation to the June-July climatology of NCEP Final Reanalysis. This modified data was used as initial conditions for the vortex initialization simulations. The construction of the vortex has been done using NCAR-AFWA tropical cyclone (TC) "bogussing" scheme \citep{davis2001ncar,fredrick2009bogussing,low2001development}. This method is not computationally expensive and follows the vorticity removal and inversion method to remove any preexisting vortex and insert a prescribed vortex in nonlinear balance. This method has been widely used as a "bogussing" scheme in the fifth-generation National Center for Atmospheric Research–Pennsylvania State University (NCAR–Penn State) Mesoscale Model (MM5) system. An updated version has been implemented in WRF \citep{fredrick2009bogussing,davis2002detection}. In particular, this method is widely utilized in enhancing the description and details of tropical cyclone initialization \citep{jian2008numerical} and in correcting the location and intensity of the cyclone
precursors \citep{ding2004simulation,kuester2008model,yang2008modeling,van2011high,wang2008comparison,komaromi2011diagnosing}. 
The scheme runs in two primary steps. The first step aimed to find the background state by removing perturbed fields associated with the regional vorticity by solving a series of Poisson's equations. In the second step, a user-defined balanced vortex of prescribed winds and moisture profiles is placed in the specified location. Since our background state is already known as climatology, we only use the vortex addition part of the scheme. For further details of method please see Appendix A-1.

\begin{table*}
\caption{Model Specifications}
\centering
\label{T1}
\begin{tabular}{lll}
\toprule
Sr.No & Attribute & Attribute Characteristic\\
\midrule
1&Model&Weather Research And Forecast Model (WRFV4.0) \\
2&Model Mode & Non-Hydrostatic\\
3&Time step for Integration &50 second\\
4&Number of Domain & Single Domain\\
5&Central point of the domain& $20^{\circ}$N, $80^{\circ}$E \\
6&Horizontal grid distance (Model Grid Resolution)& 57 km\\
7&Map Projection&Mercator\\
8&Numberof grid points &X- direction 100\\
&&Y- direction 68\\
9&Horizontal grid distribution&Arakawa C-grid\\
10&Nesting&No nesting\\
11&Vertical model level & Terrain-following
hydrostatic-pressure coordinate\\
&& (33 sigma levels up to 50 hPa)\\
12&Spatial differencing scheme&6th-order centered differencing\\
13&Initial condition&Three-dimensional real-data $(1^{\circ}\times ~1^{\circ}$ FNL)\\
14&Lateral boundary conditions&WRF specified option\\
15&Time integration&3rd-order Range-Kutta\\
15&Micro Physics & WRF Single–moment 6–class Scheme\\
15&Surface Layer & MM5 Similarity Scheme	\\
15&Planetary Boundary Layer (PBL) &Yonsei University Scheme (YSU)\\
16&Radiation& RRTMG Shortwave and Longwave Schemes \\
17&Land surface Model&Unified Noah Land Surface Model
\\
18&Cumulus parameterization schemes&New Tiedtke Scheme \\
19&Turbulence and mixing option&Evaluates 2ed order diffusion\\
20&Eddy coefficient option&Horizontal smagorinsky first order closure\\
% 21&6th order numerical diffusion&No\\
% 22&Eddy coefficient option&Horizontal smagorinsky first order closure\\
% 23&6th order numerical diffusion &No\\
% 24&W-damping&No damping\\
% 25&Base temperature&290K\\
% 26&Upper level damping&No damping\\
\bottomrule
\end{tabular}
\end{table*}

\subsection{Model Setup}

We utilize WRF Model version 4 to perform numerical experiments. Since we aim to understand mainly the large-scale interactions between MTCs and Bay of Bengal LPS, a relatively coarse horizontal grid resolution of 57 km has been used with 33 vertical hybrid sigma levels in the vertical. First, we calculate 16 years (2000-2015) of 6 hourly climatologies from 20 June to 10 July. The 2000-2015 time interval is chosen for climatology as most of the input variables are consistent in dimensions and resolutions in this time window, outside of which there are differences in the soil levels and vertical resolutions of the input data. The calculated six-hourly climatology serves as the lateral boundary conditions for the simulations. 
%$Note that we retained the diurnal cycle while calculating the climatology. In other words, no daily mean is performed to reduce the noise, which may arise due to differences in the diurnal cycle of the inner domain and climatological lateral boundary conditions if daily mean were performed.
Sea surface temperature (SST) and soil moisture were kept to their climatological value. Vortices have been added to the climatology in different locations over the Bay of Bengal to evaluate the sensitivity and to reflect the differences in locations of the Bay of Bengal LPSs. 
% \begin{equation}
% S(x,y) = S_{0}~ exp\left[\left(\frac{(x-x_{0})^2}{2\sigma_{x}^2}\right)+\left(\frac{(y-y_{0})^2}{2\sigma_{y}^2}\right)\right],    
% \end{equation}
% where $S$ is the SST perturbation, $x_{0}$, and $y_{0}$ are the longitude and latitude of the center of the SST blob. $S_0$ (=280K) is the SST value at the center of the blob, and $\sigma_{x}$ and $\sigma_{y}$ determine the smoothness and extent of the blob. 
Further, the FNL data specified lateral boundary conditions, and initial conditions were utilized for model initialization in dry and cold Bay of Bengal simulations. As far as physics is concerned, several combinations of physics schemes can be used in the WRF; however, not all combinations are tested and suitable for weather phenomena of different regions of the globe and for different cases. Since Arabian Sea MTCs share similar characteristics with topical systems \cite{ksn}, we utilize well tested combination of physics for tropical systems known as the "TROPICAL SUITE" in vortex addition experiments. Specific summary of important model details is presented in Table~\ref{T1}. The simulations are initialized with a Dolph digital filter, which runs backward and forwards for 12 hours before the actual model runs to remove any initial imbalances in the initial conditions \citep{peckham2016implementation}. In the second set of experiments, the Bay of Bengal sea surface temperature was cooled by replacing the sea surface temperature over Bay of Bengal with the cold Gaussian temperature distribution (see Appendix A2 for details) and atmosphere was dried in initial conditions over Bay of Bengal and East India (see Figure-S4). After trial and error here a slightly different combination of physics schemes is used which satisfactorily reproduces the intensity, size and location of simulated July 2020 MTC compared to observation.  In particular the for Cumulus Parameterization we use Kain–Fritsch Scheme, Kessler Scheme used for  Micro Physics, other schemes are used from "CONUS SUITE", specifically, RRTMG for radiation, Mellor–Yamada–Janjic Scheme (MYJ) for the planetary boundary layer, Eta Similarity Scheme for surface layer, and Unified Noah Land Surface Model for land surface parameterization.

\subsection{Vorticity Budget}
To understand the intensification of Arabian Sea MTCs, we use a vorticity budget \citep{holton1973introduction,raymond2011vorticity,Boos2015}. The inviscid vertical relative vorticity ($\xi$) budget equation in pressure coordinates reads,
\begin{equation}
\frac{\partial \xi}{\partial t}=-\xi\nabla.{\bf{V}}-f\nabla.{\bf{V}}-{\bf{V}}.\nabla{\xi}-{\beta}{v}-\omega\frac{\partial \xi}{\partial p}+ \left(-\frac{\partial \omega}{\partial x}\frac{\partial v}{\partial p}+\frac{\partial \omega}{\partial y}\frac{\partial u}{\partial p}\right)+ \textrm{Res}.
\label{GQ:vorbudget}
\end{equation}
In equation $\xi$ represents relative vorticity, $\beta$, meridional gradient of Coriolis parameter, $\omega$, vertical pressure velocity, $u$, zonal wind component, $v$, meridional winds, $\bf{V}$ is the total wind vector.  Here, the first two terms on the RHS represent the vorticity generation (destruction) by convergence (divergence) of horizontal winds when coupled with the relative and planetary vorticity, respectively. The third term represents vorticity advection by the horizontal wind (${\bf V})$, the fourth term is the coupling between differential planetary vorticity and meridional wind (the so-called $\beta$-term), the fifth term represents the vertical advection of relative vorticity via upward or downward motion, and the sixth term is the tilting or twisting term which is the change in vorticity due to horizontal gradients in the vertical velocity or vice versa. Finally, the last term is a residual that arises due to the parameterized physics and numerical approximations in post-processing the data. To understand the formation and intensification of MTC, each term of the vorticity equation is calculated and compared to deduce the dominant contributions to the systems' growth.

\section{Results and Discussions}

\section{Evolution of Dynamic Fields During Type 2a MTCs}

\begin{figure*}
\centering
\includegraphics[trim=0 0 0 0, clip,height = 1\textwidth,width = 0.8\textwidth, angle =90, clip]{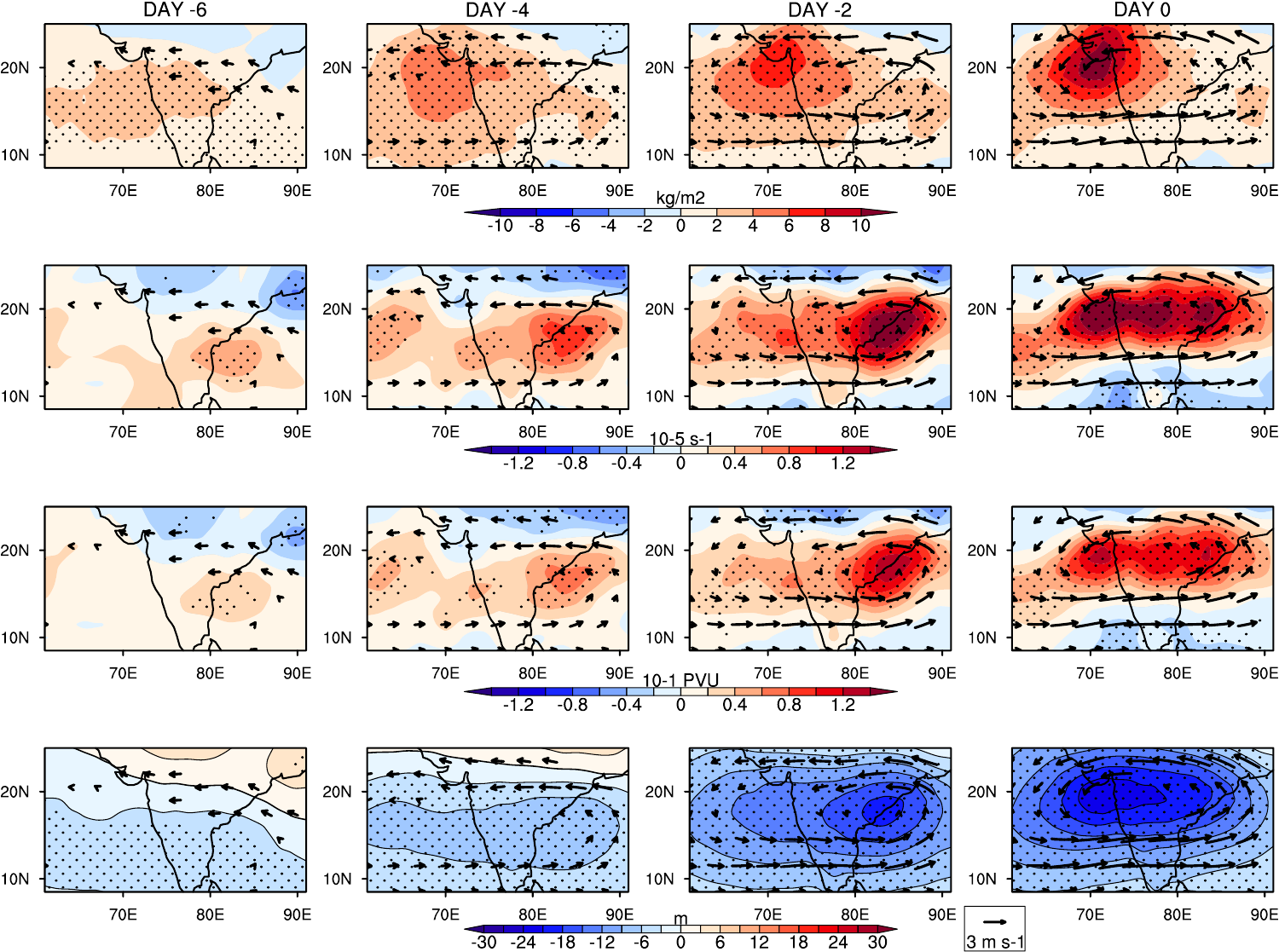}
\caption{Lag composite of 60 Type 2a systems \cite{kushwaha2023classification}, averaged over the 800-400 hPa layer and spanning from day $-6$ to Day 0. Here, Day 0 represents the day of maximum precipitation during the lifespan of the MTC. The figure presents anomalies of (a) precipitable water, (b) relative vorticity, (c) potential vorticity, and (d) geopotential height. The dotted region indicates areas where the fields significantly deviate from zero at a 0.1 significance level. Wind vectors are included only if any wind component significantly deviates from zero at a 0.1 significance level, determined using a two-tailed Student's t-test. } 
\label{fig:FIG1}
\end{figure*}

\noindent We aim to investigate Bay of Bengal systems' role in forming Arabian MTCs, specifically those that form with a preceding and coexisting LPS over the Bay of Bengal, i.e., the so-called Type 2a MTCs \citep{ksn}. We first discuss features of Type 2a systems and their connection to Bay of Bengal LPSs from reanalysis data. Plan views of Type 2a  MTC lag composites of anomalies of total precipitable water, relative vorticity,  potential vorticity (PV), geopotential height, and wind vectors are shown in Figure~\ref{fig:FIG1} in rows 1, 2, 3, and 4, respectively. Where applicable, variables are vertically averaged between 800-400 hPa to ensure that composite anomalies have a coherent vertical structure and are robust through the depth of the middle troposphere. The horizontal extent and magnitude of precipitable water anomalies (row 1; Figure~\ref{fig:FIG1}) increase over the northeast Arabian Sea from Day $-$6 to Day 0, i.e., till the day of heaviest precipitation. During this period, the wind anomaly (quivers in Figure~\ref{fig:FIG1}) is easterly and slightly northeasterly over the northern Arabian Sea, where the maximum accumulation of water vapor is observed. The easterly wind anomalies are not localized to the Arabian Sea region but appear as part of a large-scale circulation; indeed, wind anomalies originate from the Bay of Bengal and reach up to the Arabian Sea, enveloping the southern peninsula of India.

\noindent Sequentially, on Day $-$6, a weak negative height anomaly envelops the southern tip of India (row 4; Figure~\ref{fig:FIG1}), accompanying a slight positive vorticity anomaly over the southwest Bay of Bengal (around 15$^{\circ}$N, 85$^{\circ}$E; row 2, Figure~\ref{fig:FIG1}). The height anomaly over the Arabian Sea has a large-scale east-west oriented structure which is restricted to the south of 15$^{\circ}$N (row 4; Figure~\ref{fig:FIG1}). 
The Bay of Bengal height anomaly centered near 15$^{\circ}$N to 85$^{\circ}$E deepens from Day $-$6 to Day $-$4 with the intensification of associated relative vorticity and potential vorticity anomalies (rows 2 \& 3 of Figure~\ref{fig:FIG1}). Following this, an east-west shear zone forms extending from the Bay of Bengal to the Arabian Sea (rows 2 \& 3; Days $-$6,$-$2; Figure~\ref{fig:FIG1}). Within this evolving shear zone, over the Arabian Sea, as seen on Day $-$4, two significant positive PV and relative vorticity anomaly centers form --- one near the west coast of India near 15$^{\circ}$N and 72$^{\circ}$E and another around 65$^{\circ}$E and 18$^{\circ}$N. With the further intensification of the system over the Bay of Bengal (Days $-$4 to $-$2), the westward-situated Arabian Sea vorticity anomaly moves eastward, and by Day $-$2, it merges with the anomaly over the western coast of India. Note that as the height anomaly over the Bay of Bengal deepens and moves northward (Day $-$6 to Day $-$2), the trough over the Arabian Sea intensifies and becomes more localized off the coast of Maharashtra. This is reflected in the rapid increase of vorticity and PV anomalies over the northeast Arabian Sea from Day $-$2 to Day 0 (row 3; Figure~\ref{fig:FIG1}). 

\noindent Overall, the evolution of height, velocity fields, and winds suggest that anomalies over the Arabian Sea and western India are indeed linked with the Bay of Bengal system --- but, the anomalies first deepen over the Bay of Bengal, and this is followed by an increase in vorticity and PV over the Arabian Sea. The intensification of the Bay of Bengal system before that of the Arabian Sea region seems reasonable as, during the primary monsoon months, the Bay of Bengal SST and precipitable water are relatively high compared to the Arabian Sea \citep{masunaga2014free,saikranthi2019variability,saikranthi2019differences}. These conditions are favorable for the development and intensification of cyclonic systems. On the contrary, the deep convection over the Arabian Sea is usually hampered by colder SSTs due to the upwelling of subsurface cold water and further by a dry middle troposphere and a low-level inversion \citep{narayanan2004role,muraleedharan2013study,das2021characteristics}.  

\noindent The composite Type 2a MTC circulation in Figure~\ref{fig:FIG1} bears a close resemblance with the flow during rainy days over western India \citep[Figure 1f;][]{kushwaha2022classification}; specifically, both are characterized by the westward extending height depression and associated circulation which originates over Bay and ends over the Arabian Sea. Further, the structure of geopotential height at 600 hPa on rainy days \citep[Figure 1f;][]{kushwaha2022classification} and a composite of geopotential height anomaly of Type 2a MTCs in the fourth row of Figure~\ref{fig:FIG1} show a non-circular Bay of Bengal system with a westward extending trough. This trough is remarkably similar to the steady-state Rossby response to off-equatorial heating \citep{gill1980some,goswami1987mechanism}. Thus, the apparent atmospheric link between the Arabian Sea and the Bay of Bengal basins may be mediated by the Gill response \citep{gill1980some} to off-equatorial diabatic heating over the Bay of Bengal and East India.  
Previous studies had noted the remote influence of Bay of Bengal heating on the western Indian circulation, wherein when heating was prescribed over the Bay of Bengal, cyclonic vorticity and precipitation were enhanced over western India and the northeast Arabian Sea, and this was argued to be due to a Rossby wave response to the heating \citep{xie2006role,choudhury2018phenomenological}.
More broadly, remote induction of synoptic scale systems downstream of a mature tropical cyclone has been observed in various ocean basins \citep{krouse2010observational,schenkel2016climatology,schenkel2017multiple}, including cyclogenesis in Rossby waves radiated from a parent cyclone \citep{krouse2008wavelength}.

%=================================
%=================================

\begin{figure*}
\centering
\includegraphics[trim=0 0 0 0, clip,height = 1\textwidth,width = 0.7\textwidth, angle =90, clip]{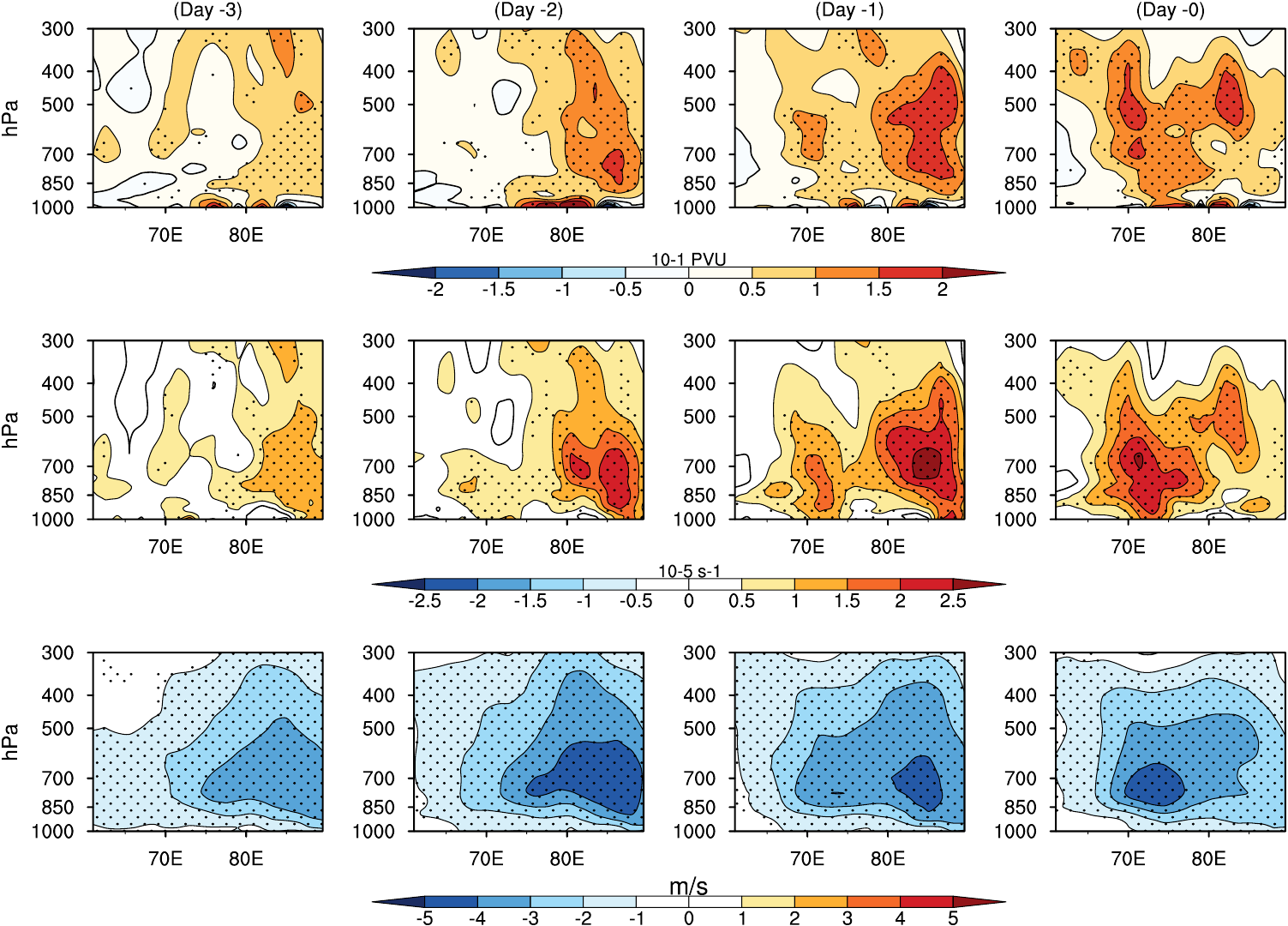}
\caption{Cross section of lag composite of 60 Type 2a systems \cite{kushwaha2023classification}, from Day $-3$ to Day 0. Here, Day 0 represents the day of maximum precipitation during the lifespan of the MTC. Row 1: Potential vorticity anomaly, Row 2: Relative vorticity anomaly, Row 3: Zonal wind anomaly, %Row 4: Meridional wind anomaly.
Shading denotes significant regions at a 0.1 significance level under a two-tailed student-t-test. The cross-section of vorticity and PV is at $18.5^{\circ} $N; of wind, component is at $22^{\circ}$N.}
\label{fig:FIG2}
\end{figure*}

 \noindent Up to now, we focused on the horizontal evolution of the system; to understand the vertical structure evolution dynamical fields during the Type 2a system formation, vertical-horizontal cross sections of composite anomalies of relative vorticity, PV, zonal winds of 60 Type-2a MTCs are shown in Figure~\ref{fig:FIG2}. Before Day $-$4 (not shown), the relative and potential vorticity are disorganized and weak in much of the troposphere over western India. However, as the low-level PV and relative vorticity spin-up over the Bay of Bengal, the corresponding mid-level entities increase in an organized and statistically significant manner over the Arabian Sea (Figure~\ref{fig:FIG2}; row 1 and row 2). At later stages (Day $-$1 to 0), the vorticity and PV intensify at the lower levels, too, reflecting the increasing presence of deep convection in the system \citep{murthy2019understanding,russell2020potential,ksn}, or a transition of the MTC into a lower troposphere cyclone phase \citep[LTC-phase;][]{ksn}. 
 
\begin{figure*}
\centering
\includegraphics[trim=0 0 0 0, clip,height = 0.7\textwidth,width = 0.5\textwidth, angle =90, clip]{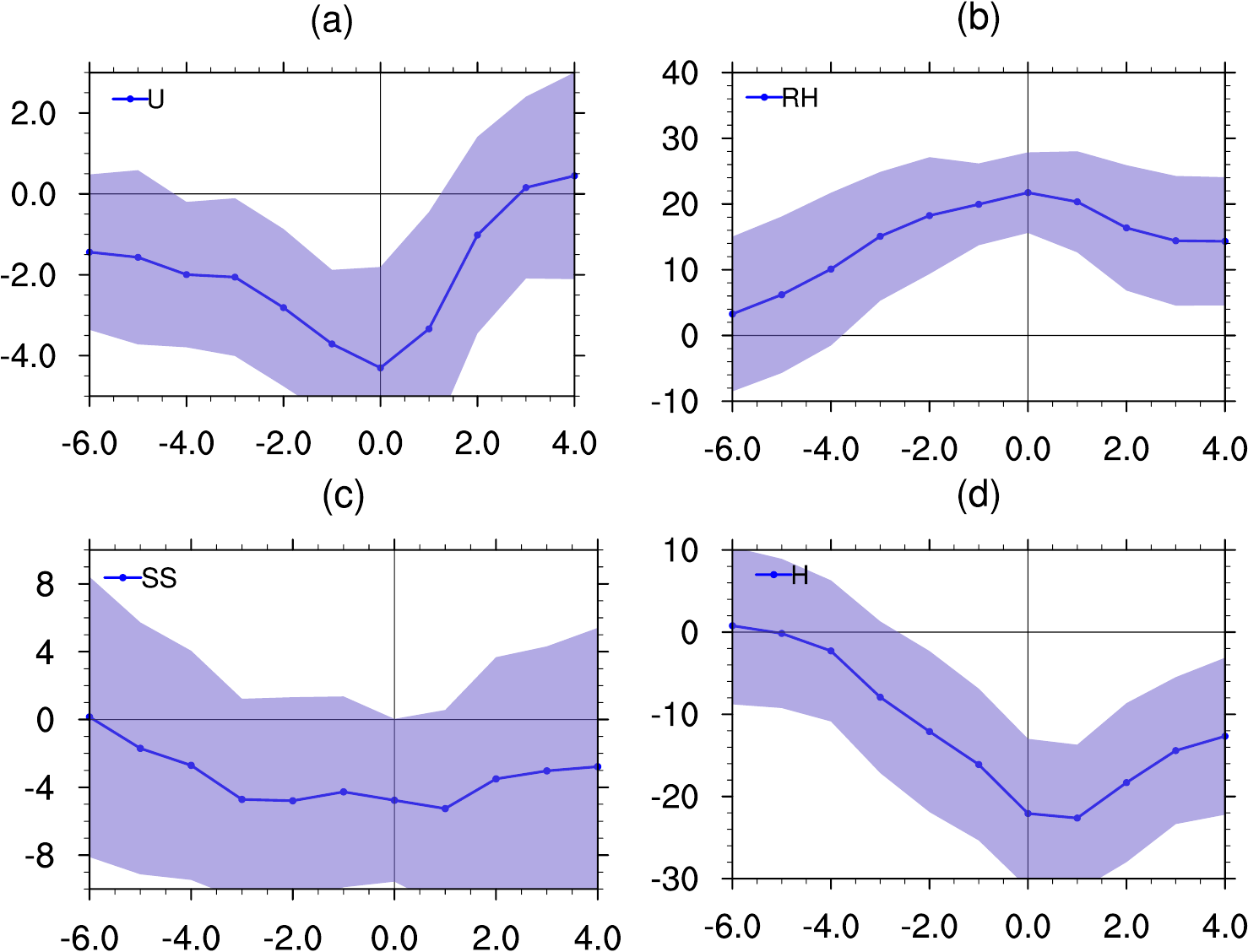}
\caption{The lag composite time series of 60 Type 2a systems \cite{kushwaha2023classification} at $22^{\circ}$N and $72^{\circ}$E spanning from day $-6$ to Day 0. Here, Day 0 represents the day of maximum precipitation during the lifespan of the MTC. The panels illustrate: (a) Zonal wind anomalies ($ms^{-1}$) averaged between 600-500 hPa, (b) Relative humidity anomalies averaged in the 600-500 hPa layer, (c) Static stability parameter mean in the 700-800 hPa layer $(10^{-5}$ K/Pa), and (d) Height anomaly (m) at 500 hPa. Shadings represent the widths of one standard deviation.} 
\label{fig:FIG3}
%\label{fig:FIG2}
\end{figure*}

\noindent The zonal wind (third row; Figure~\ref{fig:FIG2}) shows a significant increase of an easterly wind anomaly in the middle troposphere from Day $-$3 to Day $-$1 over western India, which extends up to 300 hPa with a maximum around 700 hPa. Thus, it appears that the anomalous easterlies in the middle troposphere are enhanced by the system in the Bay of Bengal. These enhanced easterlies help prevent dry air mixing from the north and west and decrease the strength of inversion by reducing elevated warm air advection \cite{dwivedi2021variability,narayanan2004role}. In turn, these favorable conditions allow for the development of deep convection and help further moisten the free troposphere over the Arabian Sea. The increase in strength of the anomalous easterlies and their influence is more clearly evident in Figure~\ref{fig:FIG3}. From Day $-$6 to Day 0, middle tropospheric easterlies become stronger (Figure~\ref{fig:FIG3}a), and concomitantly, relative humidity increases over western India (Figure~\ref{fig:FIG3}b). The expected reduction in low-level static stability or the reduction in strength of the low-level inversion is also observed (Figure~\ref{fig:FIG3}c). A weaker inversion allows for an increase in the middle troposphere moisture which favors deep convection \cite{raymond2015balanced} and the further spin-up of an MTC and deepening the height anomaly as seen in Figure~\ref{fig:FIG3}d.

\noindent To a first approximation, we hypothesize that sufficient localized latent heating in the atmosphere above the Bay of Bengal and East India results in a off equatorial Gill-type response consisting of a westward extending zonal trough that reaches up to the Arabian Sea starting from Bay of Bengal \citep{goswami1987mechanism,gill1980some,xie2006role,choudhury2018phenomenological}. %{\textcolor{red}{Ravi Sir suggested one paper "Arindam et al 2009" which suggest the link of Arabian sea winds with the Bay of Bengal heating., however, i 
%am unable to locate that paper}} 
Given its rotational character, this Rossby gyre is characterized by enhanced horizontal shear and cyclonic vorticity over western India. A similar response to realistic diabatic heating in the middle troposphere has been noted in recent general circulation model experiments \citep{xie2006role,choudhury2018phenomenological}. Moreover, the enhanced horizontal shear associated with the Rossby gyre can be critical for the growth of the Arabian Sea system by barotropic instability \citep{goswami1980role,goswami1987mechanism}; in fact, moist barotropic instability has been identified as a possible source of energy for the formation of monsoon lows and MTCs \citep{goswami1980role,diaz2019barotropic,diaz2019monsoon}.
In addition to the shear and vorticity, the enhanced easterlies north of the northeast Arabian Sea and western India prevent the inflow of desert air, thus weakening the climatological inversion and raising the possibility of convection that can moisten the free troposphere. 
The altered dynamic and thermodynamic conditions, forced by LPS-induced heating in the Bay of Bengal, in turn, favor the formation of MTCs over the northeast Arabian Sea and western India.
%=================================
% %=================================

%=================================
% %=================================

%=============================================
%           Numerical Experiments 
%=============================================

%\section{Numerical Simulation of a MTC}

\section{Bogus LPS over the Bay of Bengal}
\begin{figure}    
\centering
%\vspace{-1.5cm}

\includegraphics[trim=0 0 0 0, clip, height = 0.7\textwidth, width = 0.5\textwidth, angle = 90, clip]{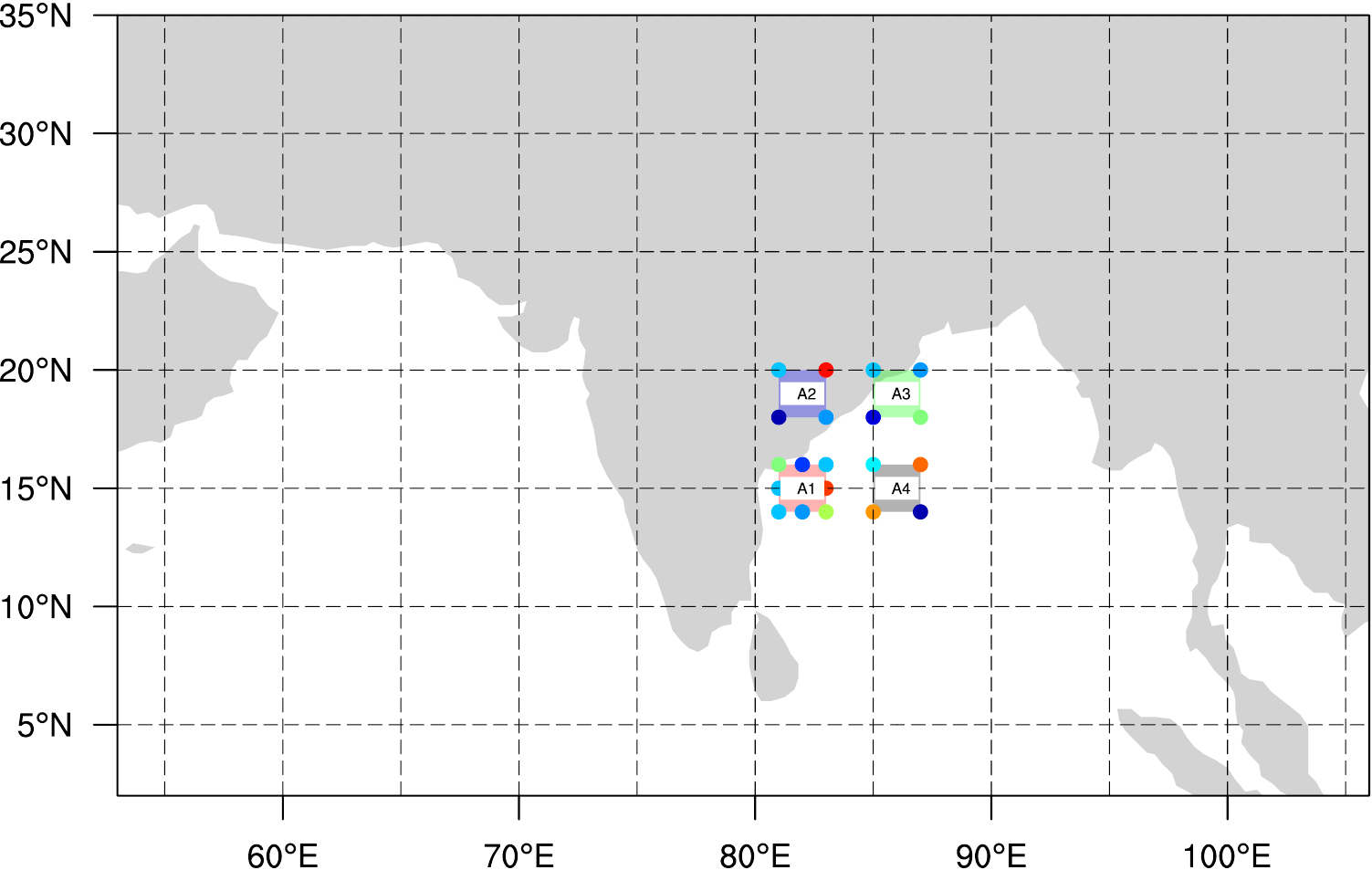}
\caption{ The model domain, with locations marked for 21 bogus vortices divided into four sets: A1, A2, A3, and A4. Group A1 comprises nine members, while groups A2 to A4 consist of four members each.}
\label{fig:FIG4}
\end{figure}

\noindent To verify the hypotheses suggesting that the systems in the Bay of Bengal play a significant role in creating a favorable environment over the Arabian Sea, and subsequently contribute to the genesis MTCs, we now conduct series of numerical experiments utilizing the Weather Research and Forecasting (WRF) model. The details of the model, its configuration, setup, and the bogus vortex technique have been described in the Methods section. The model domain is shown in Figure \ref{fig:FIG4}.
The domain is chosen to be large enough to include several important components of the monsoon, such as the Somali Jet, the monsoon trough, the heat low, Bay of Bengal, and, the Arabian Sea. In the first set of experiments, twenty-one bogus vortices (or monsoon lows) are added over different locations in the Bay of Bengal (Figure~\ref{fig:FIG4}). 
More attention has been given to region A1 with nine bogus vortices (their coordinates are: 14N, 81E; 15N, 81E; 16N, 81E; 14N, 82E; 15N, 82E; 16N, 82E; 14N, 83E; 15N, 83E; 16N, 83E) as this is where the first signs of an LPS appear on Day $-$6 in the composite plan views of Type 2a formation (Figure~\ref{fig:FIG1}). The rest of the 12 members are added to other locations in regions A2 to A4\footnote{
The location of other ensemble members in regions A2 to A4 has been chosen to check the sensitivity of MTC induction to different locations of lows in the Bay of Bengal. Note that the intensity and size of the vortex are similar to the IMD data and observed sizes of Bay of Bengal LPSs \citep{Hunt}, respectively. Weak and small vortices are advected by mean flow and take longer to intensify; hence these particular values were selected by trial and error.}. Specifically, the coordinates in region A2 are: 20N, 81E; 20N, 83E; 20N, 85N; 20N, 87E, region A3  are: 18N, 87E; 16N, 87N; 14N, 87E; 18N, 81E, and region A4 are 18N, 83E; 18N, 85E; 14N, 85E; 16N, 85E. Note that we use slightly reduced sizes ($r_{m} =200$ km against $r_{m} =350$ km) in A1 to overcome the difficulties of model instabilities due to the effects of the topography of Himalayan regions at higher latitudes during dynamical adjustment. Further, $v_{max} =14$ ms$^{-1}$ or 27 kt has been chosen, which is the strength of a monsoon depression defined by the IMD.  

\begin{figure}    
\centering
\includegraphics[trim=0 0 0 0, clip, height = 0.9\textwidth, width = 0.8\textwidth, angle = 90, clip]{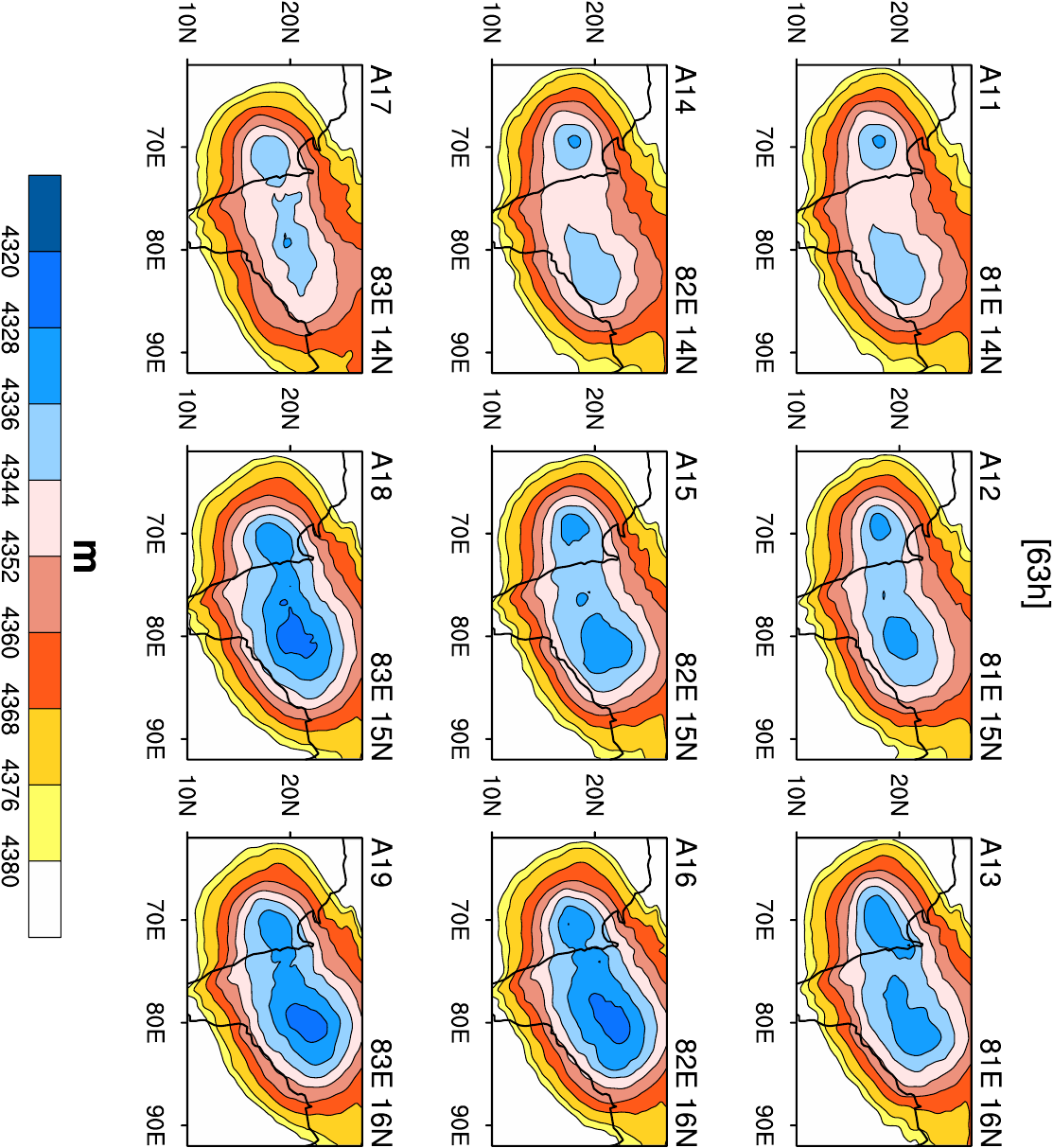}
\caption{Geopotential height at 600 hPa of nine members (A11-A19) of group A1 after 63 hours of simulation. Each subplot represents the height fields of respective vortex located at nine different positions. }
\label{fig:FIG7}
\end{figure}

\noindent The 600 hPa geopotential height after 24 hours of simulation for nine members of group A1 is shown in Figure~S1. 
As expected, all ensemble members show lows in the locations where bogus vortices were added along with a trough extending from the Bay of Bengal up to the Arabian Sea and western India. Apart from the trough, there is no signature of MTC existence over the Arabian Sea. However, the heights at 600 hPa after 60 hours of model integration (Figure~\ref{fig:FIG7}) suggest the existence of MTCs, as evident from closed height contours centered around 70$^{\circ}$E and 18$^{\circ}$N in all the nine ensemble members region A1. Figure~S3 depicts the entire formation cycle of an MTC in a single ensemble member, where a bogus vortex is introduced at 82$^\circ$E and 15$^\circ$N (center of region A1). It is important to note that similar evolution is observed in other ensemble members. The time evolution clearly illustrates that as the Bay of Bengal system progresses northwards, the associated height perturbations extend westward, covering parts of western India and the Arabian Sea as a westward-extending zonal trough from 24 to 36 hours. Within the western end of the trough, which extends over portions of the Arabian Sea, a closed vortex emerges after about 48 hours of simulation and reaches maturity as a closed vortex at approximately 60 hours of simulation. 
Further, this zonal trough connects the induced MTC with the bogus Bay of Bengal vortex. Both systems encircle each other similar to Type 2a composites \citep[Figure \ref{fig:FIG1};][]{kushwaha2022classification} and with specific MTC observations \citep{miller1968iioe}. Indeed, a close inspection of the location of the trough and MTC formation site in Figure S1, S3 and Figure \ref{fig:FIG7}, respectively, suggest that MTC formation occurred in the middle tropospheric trough, which was induced by the bogus Bay of Bengal vortex.

% {\textcolor{blue}{To further investigate the influence of Bay of Bengal LPSs on the genesis of Arabian sea MTCs, we conducted a free simulation using climatological initial conditions, without the introduction of any vortex. Figure S7 illustrates the time evolution of geopotential height at 600 hPa. Initially, a weak high depression and zonal trough, associated with the climatological high fields, are observed for the first 48 hours. Subsequently, the Bay of Bengal trough gradually intensifies, leading to deepening height anomalies and the emergence of a well-developed monsoon low at approximately 60 hours into the simulation. Notably, as the low-pressure system intensifies, the associated trough extends towards the Arabian Sea and western India. Around 96-108 hours into the simulation, MTCs begin to develop within this trough. These findings provide further evidence that the formation of MTCs is contingent upon the development and intensification of the monsoon low over the Bay of Bengal. 
% %It is noteworthy that the onset of MTCs in the free simulation is delayed by approximately 2-3 days compared to the vortex addition simulations, reflecting the time required for the natural formation of the monsoon low-pressure system from the climatological monsoon trough over the Bay of Bengal. 
% These results underscore the pivotal role of the Bay of Bengal system in establishing a favorable environment for the genesis of MTCs over the Arabian Sea.}} 

\begin{figure}    
\centering
\includegraphics[trim=0 0 0 0, clip, height = 0.9\textwidth, width = 0.8\textwidth, angle = 90, clip]{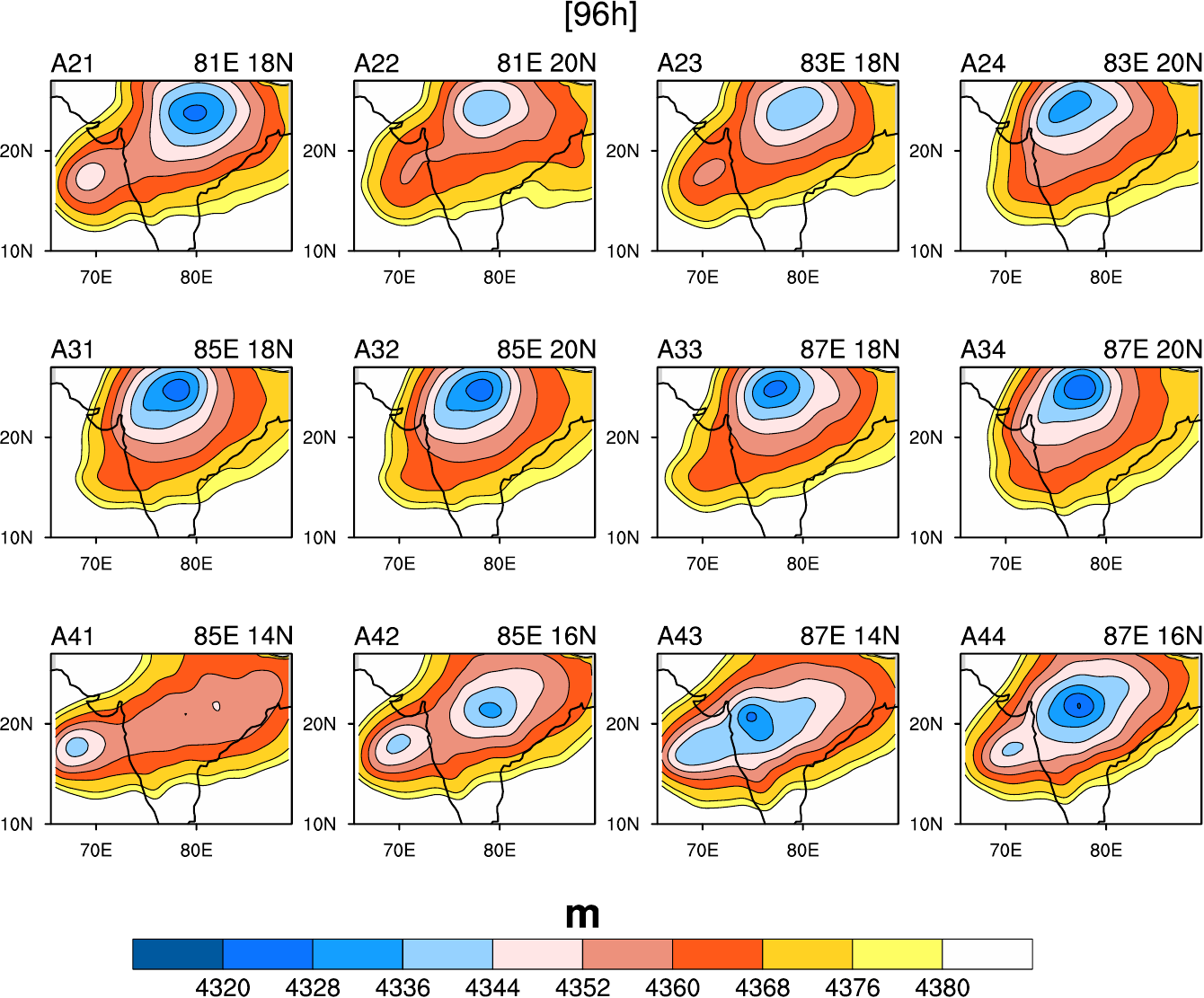}
\caption{ The geopotential height at 600 hPa for twelve members of group A2 (row 1), group A3 (row 2), and group A4 (row 3) after 96 hours of simulation, respectively. Each subplot in the rows represents the height fields of the respective vortex located at four different positions.}
\label{fig:FIG9}
\end{figure}

\noindent To understand the sensitivity of MTC genesis to different locations of Bay of Bengal LPSs, other ensemble members were added over A2, A3, and A4 regions shown in Figure \ref{fig:FIG4}. The 600 hPa geopotential surface after 24 hours of model integration of group A2 to A4 is shown in Figure S2. 
During the first 24 hours of simulation, similar to group A1, we see only a westward extending trough up to the Arabian Sea and no closed vortex or existence of MTC. However, after 96 hours of model integration, the signature of MTC becomes evident over the Arabian Sea and western India in group A2 (first row; Figure~\ref{fig:FIG9}) and A4 (third row; Figure~\ref{fig:FIG9}). For the members of set A3 (second row; Figure~\ref{fig:FIG9}), we only observe a trough formation, which does not develop into a closed-form vortex or MTC for up to 96 hours of model integration. 

\noindent This aligns with the findings of \citet{kushwaha2023classification}, who demonstrated that the majority of {\it in-situ} MTCs are associated with Bay of Bengal LPSs that have a relatively southern track, while majority of LPSs taking relatively northern tracks are not always connected to rainy MTCs over the Arabian Sea. These results strongly support the hypothesis that a robust dynamic link exists between Bay of Bengal LPSs with relatively southern tracks and the genesis of Type 2a MTCs over the Arabian Sea and western India. Remarkably, when a bogus vortex is introduced to the climatological conditions in favorable locations over the Bay of Bengal, an MTC is triggered over western India or the Arabian Sea within a timeframe of 60-100 hours (2.5 to 4 days), clearly highlighting the significant influence of the Bay of Bengal system.

\subsection{Vorticity Budget}

\begin{figure}    
\centering
\includegraphics[trim=0 0 0 0, clip, height = 0.4\textwidth, width = 1\textwidth, angle = 0, clip]{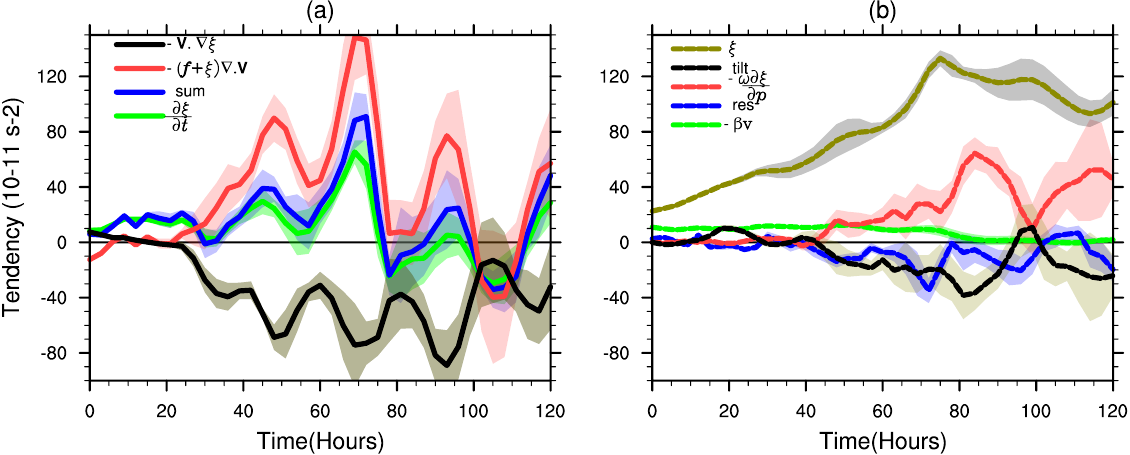}

\caption{\textcolor{black}{Composite derived time series of various terms of vorticity budget of ensemble A1, averaged over the MTC region (15-20$^{\circ}$N, 65-72$^{\circ}$E) between 500--600 hPa. The time axis corresponds to the simulation time. The units of vorticity tendency terms are $10^{-11}$$s^{-2}$. Note that, for comparison, the relative vorticity, $\xi$, in Figure 7b is multiplied with $2\times10^{-6}$ hence units are $0.5\times10^{-6}$ $s^{-1}$. Shading represents one standard deviation among A1 ensemble members. In (b), legend "res" represent residual in vorticity budget while "tilt" stands for tilting term.}}
\label{fig:FIG10}
\end{figure}

The genesis of MTCs in all the ensemble members from region A1 immediately leads to the question of the dominant terms of the vorticity budget that contribute to the growth of the cyclonic system over the Arabian Sea and western India. 
The composite time series of the various terms in the vorticity budget (details of the budget are in the Methods section) averaged over the MTC region (15-20$^{\circ}$N, 65-72$^{\circ}$E) between 500-600 hPa is shown in Figure \ref{fig:FIG10}. 
The observed vorticity tendency (Figure \ref{fig:FIG10}a green curve) and the sum of all terms on the right side of the vorticity budget (Equation \ref{GQ:vorbudget}, Figure \ref{fig:FIG10}a blue curve) match quite well, suggesting a well-balanced budget. In particular, during the first 24 hours of simulation, the positive vorticity tendency is mainly accounted for by the advection (mainly $\beta$) and tilting terms (Figure \ref{fig:FIG10}a and b).
Notably, the stretching term does not contribute during the initial 24 hours. The vorticity tendency increases rapidly after 24 hours of simulation, reaching its maximum of around 70 hours. In contrast to the first 24 hours, the vortex stretching term accounts for most of the intensification from 24 to 72 hours (Figure \ref{fig:FIG10}a, red). Further, the coupling of relative vorticity with divergence accounts for a slightly larger portion of the stretching than the coupling of planetary vorticity with divergence (not shown). On the other hand, during the same period, the horizontal advection of vorticity (Figure \ref{fig:FIG10}a black) acts as a major sink and opposes vortex stretching. Though the $\beta$-term (Figure \ref{fig:FIG10}b, green) dominates during the first 24 hours and continues to act acts as a mild source of vorticity throughout the formation of MTC, it is relatively small compared to the stretching term during the rapid intensification period (24-70 hours). The vertical advection of vorticity (Figure \ref{fig:FIG10}b, red) also acts as a source of vorticity; however, it peaks after the maximum intensification, suggesting that it does not contribute significantly during the growth phase of MTC; also, as it peaks after the intensification of the middle-level maximum, this suggests a shift of MTC vorticity maximum to the lower levels \citep[i.e., the LTC phase,][]{ksn} leading to the positive vertical advection down the gradient of relative vorticity in an environment of upward velocity. 

\begin{figure}   
\vspace{-0.3cm}
\centering
\includegraphics[trim=0 0 0 0, clip, height = 0.9\textwidth, width = 0.8
\textwidth, angle = 90, clip]{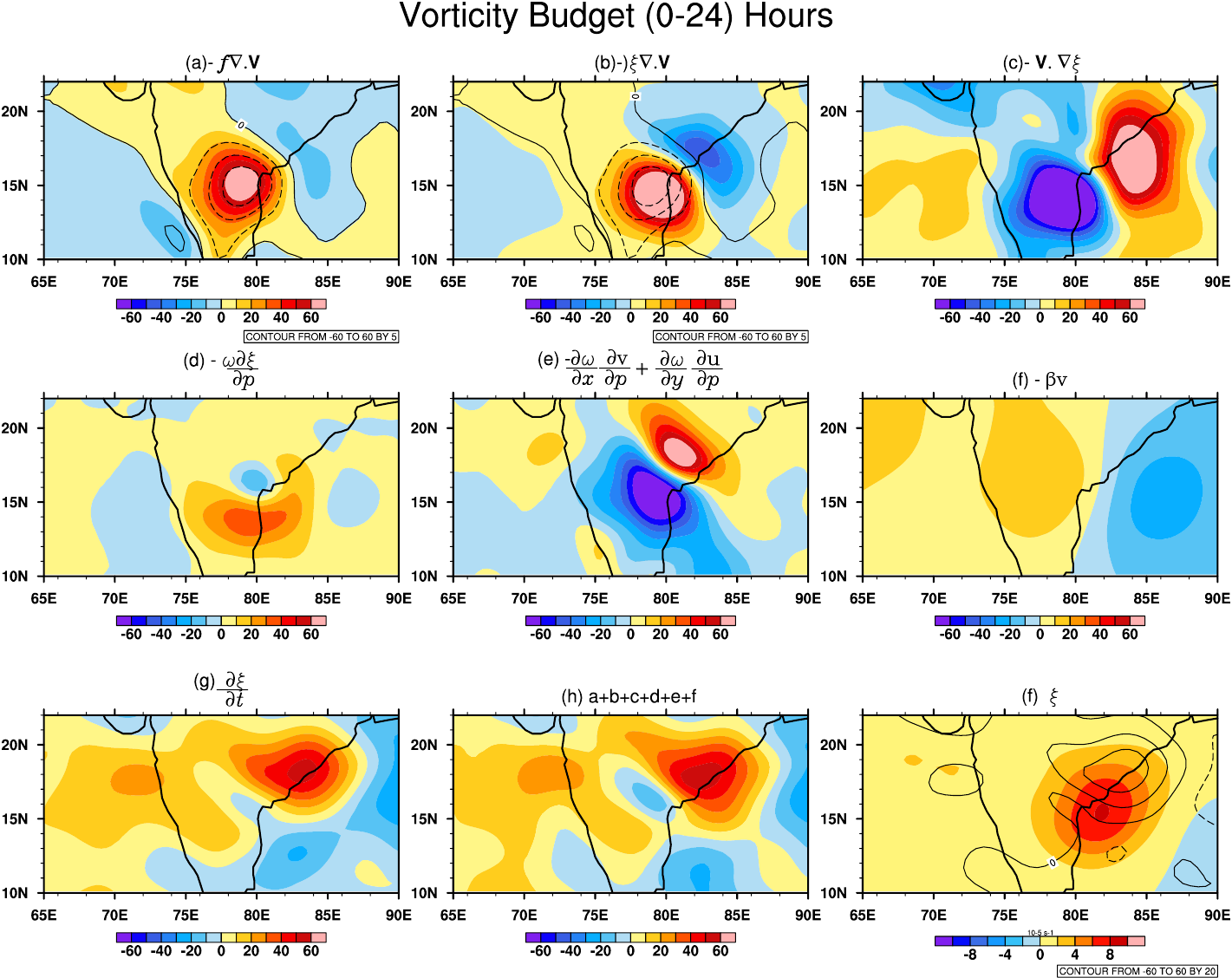}

\caption{Plan view of various terms in the vorticity budget of ensemble A1, averaged between 500-600 hPa and averaged over first 24 hours of simulation; dashed and solid contours represent convergence and divergence respectively in (a) and (b); dashed and solid contours in (f) are negative and positive $\frac{\partial \xi}{\partial t}$ respectively. Symbols have their usual meaning.} 
\label{fig:FIG11}

\end{figure}

\noindent To get a spatial feel for the various terms, a plan view of the vorticity budget during the initial 24 hours is shown in Figure~\ref{fig:FIG11}; the vorticity tendency ($\frac{\partial \xi}{\partial t}$;
Figure~\ref{fig:FIG11}g) matches the sum of all terms on the right side of the vorticity equation (Figure~\ref{fig:FIG11}h), suggesting an approximate closure of the vorticity budget. Two centers of positive vorticity tendency are observed; a strong one over the Bay of Bengal and East India --- related to the intensification of Bogus LPS, and another relatively weaker maximum over the northeast Arabian Sea off the coast of Mumbai linked to the incipient MTC. Consistent with the times series in Figure~\ref{fig:FIG10}, during the first 24 hours, the total positive vorticity tendency over the Arabian Sea (Figure~\ref{fig:FIG11}h) mainly results from the tilting term (Figure~\ref{fig:FIG11}e), 
the $\beta$-term (Figure~\ref{fig:FIG11}f) and advection of relative vorticity (Figure~\ref{fig:FIG11}c). The $\beta$-term (Figure~\ref{fig:FIG11}f) shows a broad positive tendency over the Arabian Sea in regions of southerly winds in the western sector of the vortex or LPS. The stretching term (Figure~\ref{fig:FIG11}a,b) remains positive over western India; however, it contributes less relative to tilting (Figure~\ref{fig:FIG11}e) and advection of absolute vorticity (Figure~\ref{fig:FIG11}c+f). Notably, vertical advection (Figure~\ref{fig:FIG11}d) does not appear to be a major vorticity source during the first 24 hours over the Arabian Sea. Thus, during the initial phase of the genesis of MTC, the advection of absolute vorticity and tilting of horizontal vorticity vector by the meridional gradients of the vertical motion explains almost the entire geographical distribution of the vorticity tendency.

\begin{figure}   
\vspace{-0.3cm}
\centering
\includegraphics[trim=0 0 0 0, clip, height = 0.9\textwidth, width = 0.8\textwidth, angle = 90, clip]{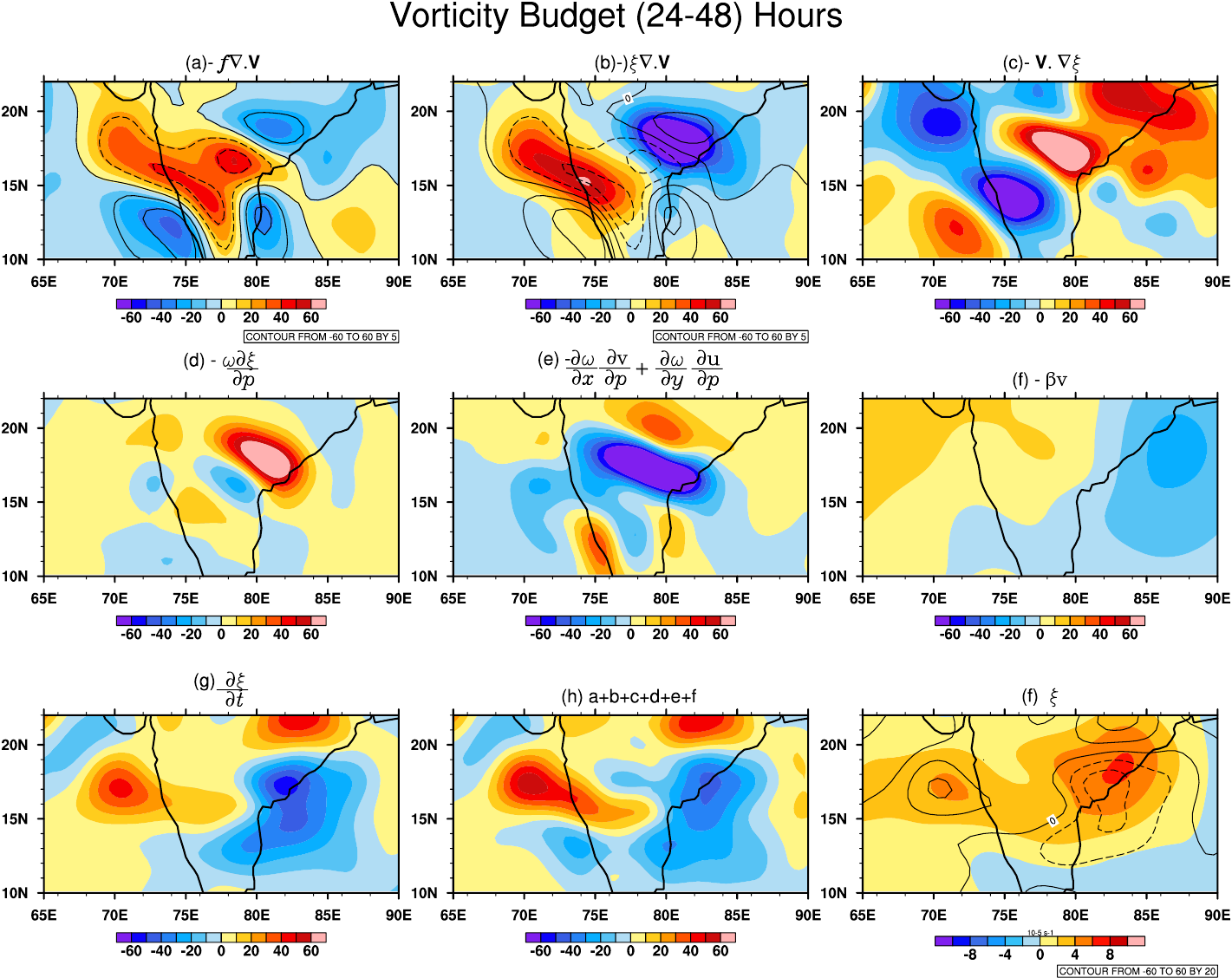}

\caption{Plan view of various terms in the vorticity budget of ensemble A1, averaged between 500-600 hPa and averaged over 24-48 hours of simulation; dashed and solid contours represent convergence nad divergence respectively in (a) and (b); dashed and solid contours in (f) are negative and positive $\frac{\partial \xi}{\partial t}$ respectively. Symbols have their usual meaning.} 
\label{fig:FIG12}

\end{figure}

\noindent The plan view of the vorticity budget during 24-48 hours of model integration is shown in Figure \ref{fig:FIG12}. The pattern of observed vorticity tendency (Figure \ref{fig:FIG12}g) and the sum of the right-hand side of the vorticity equation (Figure~\ref{fig:FIG12}h) show a similar magnitude and horizontal distribution, again suggesting a sufficiently balanced budget. In contrast to the budget during the first 24 hours, the total vorticity tendency during 24-48 hours is almost entirely explained by the total stretching term (Figure \ref{fig:FIG12} a,b). However, again, note that though the total tendency follows stretching, advection acts to cancel it and reduces its effectiveness in the intensification process. The opposing nature of stretching and advection, and their almost cancellation, has also been observed in the movement of monsoon depressions \citep{boos2015adiabatic}. The $\beta$-term (Figure \ref{fig:FIG12}f) acts like a source; however, its magnitude is much lesser than the stretching term. In contrast to the first 24 hours, here, the advection of relative vorticity by horizontal winds (Figure \ref{fig:FIG12}c) and the tilting term (Figure \ref{fig:FIG12}e) act to damp the vorticity tendency. The total vorticity tendency (Figure \ref{fig:FIG12}f, lines) and regions of positive absolute vorticity maximum (Figure \ref{fig:FIG12}f, colors) almost overlap, suggesting that the vorticity tendency primarily contributes to the intensification, and not much to the motion of the incipient MTCs --- in accord with the observations of quasi-stationary nature of Type 2a MTCs \citep{carr1977mid,ksn,kushwaha2022classification}.
To a large extent, these characteristics are consistent with the vorticity budget of July 1963 MTC, which showed stretching as a major source, followed by vertical advection and $\beta$-term, while advection acted as the main sink of vorticity \citep{carr1977mid}.
Overall, the vorticity budget suggests that the advection of absolute vorticity and tilting initially provide a positive vorticity environment for MTC growth; however, during the rapid intensification phase, the vorticity stretching dominates the MTC intensification.

\begin{figure}   
\vspace{0.3cm}
\centering
\includegraphics[trim=0 0 0 0, clip, height = 0.5\textwidth, width = 0.65\textwidth, angle = 0, clip]{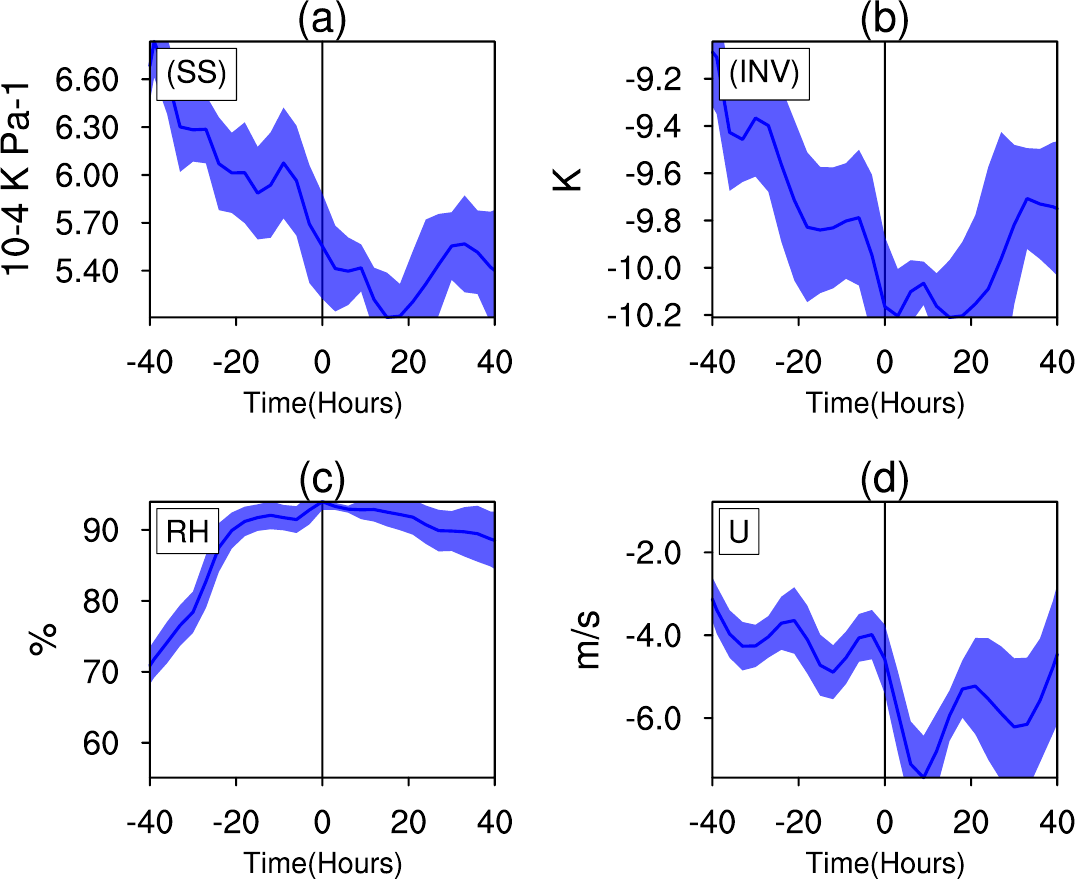}

\caption{\textcolor{black}{The composite time series, averaged over the MTC region (15--20$^{\circ}$N, 65--72$^{\circ}$E), with zero hours corresponding to the time of maximum intensification, approximately 70 hours into the simulation. The panels display: (a) the static stability parameter at 800 hPa, (b) $T_{900}-T_{700}$, representing the strength of the inversion (INV), (c) the mean relative humidity averaged over the 500-600 hPa layer, and (d) the zonal wind averaged over the 500-600 hPa layer. The shaded blue region represents the width of one standard deviation among ensemble members.} }
\label{fig:FIG_MOIST}

\end{figure}

\noindent Further, as was shown in Figure \ref{fig:FIG2}, the anomalous easterly winds from the Bay of Bengal converge over the Arabian Sea. These middle-level anomalous easterlies reduce desert air intrusions from the west and north, weaken the inversion layer, create favorable conditions for convection, and help moisten the middle troposphere. This important role of the easterlies is confirmed in the model runs for group A1 in Figure \ref{fig:FIG_MOIST}.
In particular, Figure~\ref{fig:FIG_MOIST}a and b clearly show that from -40 hours to the rapid intensification phase (i.e., before hour 0), static stability (Figure \ref{fig:FIG_MOIST}a) and strength of inversion (Figure \ref{fig:FIG_MOIST}b) continue to decrease in the lower troposphere. At the same time, the relative humidity increases (Figure \ref{fig:FIG_MOIST}c), reaching near saturation just before the maximum intensification. Similar to the observations (Figure \ref{fig:FIG3}), both simulated static stability parameter and humidity follow trends of the middle troposphere easterlies over western India (Figure~\ref{fig:FIG_MOIST}d). Moreover, the increase of middle troposphere humidity also suggests the contribution of convection in the vortex stretching. Essentially Figure \ref{fig:FIG2}, \ref{fig:FIG3}  and Figure \ref{fig:FIG_MOIST} suggest that the middle troposphere easterlies enhanced by the Bay of Bengal system eventually alter the thermal profile over the Arabian Sea and western India such that it destabilizes the lower troposphere, reduces dry air mixing from west and northwest and allows the moistening of the middle troposphere. This provides the fertile ground for deep convection, convergence, and vortex stretching and leads to MTC genesis. These are precisely the features observed during the formation of July 1963 MTC \citep{miller1968iioe}.

\subsection{Mechanism Denial Experiment}

To further confirm the role of the Bay of Bengal systems in the formation of Arabian Sea MTCs, we now consider a real instance of July 2020 MTC as a test case where the Bay of Bengal LPS preceded the formation of Arabian Sea MTC (i.e., a Type 2a member of \citet{kushwaha2023classification}). The time evolution of geopotential height at 600 hPa of four control ensembles initialized on 1 July 00, 06, 12, and 18 UTC 2020 are shown in Figure S5~%\ref{fig:FIG13} 
from Row 1 to Row 4, respectively. After 72 hours of simulation (Figure-S5, Column 1), height anomalies over the Bay of Bengal deepen, and a trough appears over the Arabian Sea and western India. With the further intensification of the Bay of Bengal LPS, the Arabian Sea trough deepens from 72-120 hours. After 120 hours of simulation, an MTC develops in all the ensemble members within the middle troposphere trough around 20$^{\circ}$N and $72^{\circ}$ E. This MTC intensifies and becomes an isolated vortex after about 120-130 hours of simulation. Thus, the model successfully replicates the July 2020 Type 2a MTC formation. 

\begin{figure}   
\vspace{-0.3cm}
\centering
\includegraphics[trim=0 0 0 0, clip, height = 0.75\textwidth, width = 0.9\textwidth, angle = 0, clip]{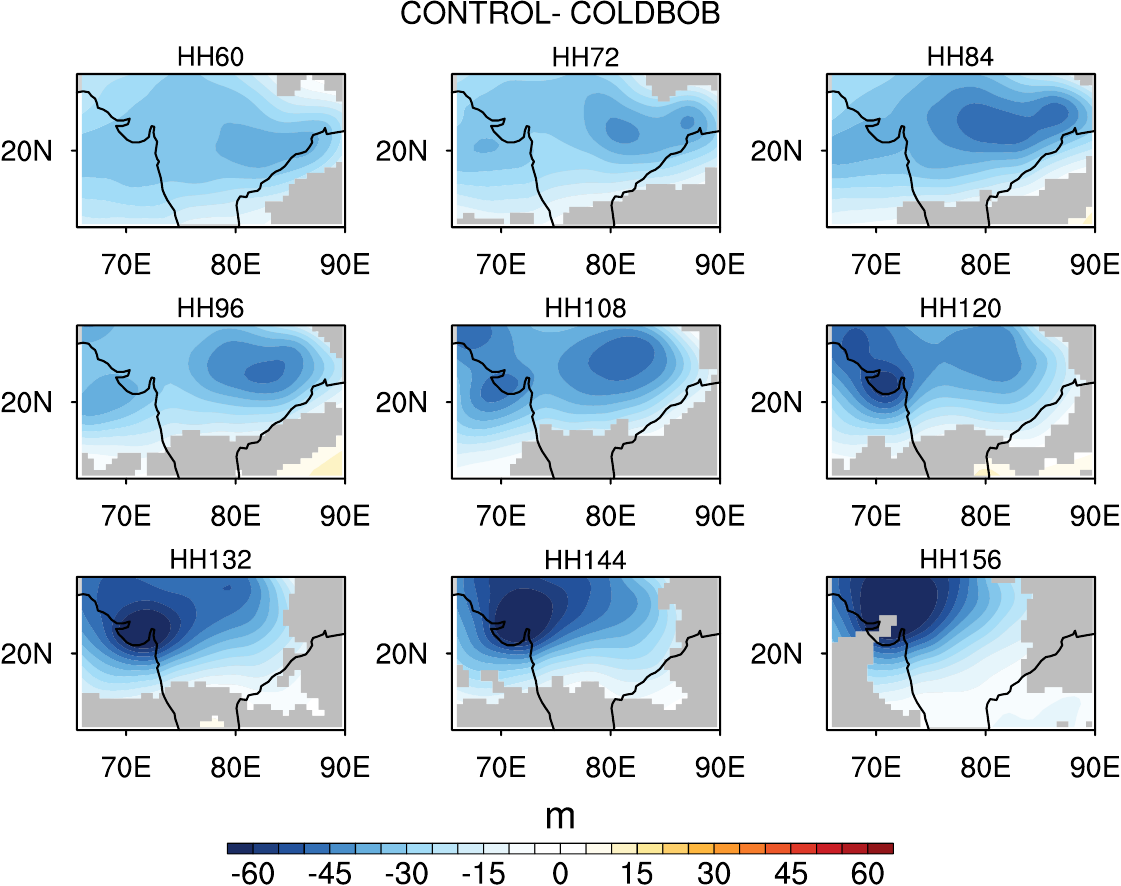}

\caption{\textcolor{black}{Simulated composite ensemble difference of geopotential height at 600 hPa between the control ensemble members and the cold-dry Bay of Bengal  case (control - cold and dry Bay of Bengal) for the July 2020 MTC. The ensemble members are initialized on 1 July at 00, 06, 12, and 18 UTC, respectively. The figures depict the simulation results starting from 60 hours (HH60) to 156 hours (HH156). Only fields that significantly differ from zero at a 99\% confidence level, determined using a two-tailed Student's t-test, are shown. Otherwise, the fields are masked. The composite is based on the common UTC dates of all members, with the 1 July 2020, 18 UTC ensemble member serving as the reference.}}
\label{fig:FIG13}

\end{figure}

\noindent We now consider a mechanism denial experiment, i.e., we check whether the MTC forms in the absence of the Bay of Bengal LPS. The suppression of the Bay of Bengal LPS is achieved by cooling and drying the Bay of Bengal, as shown in Figure S4. %\ref{fig:FIG14} 
With unfavorable conditions over the Bay of Bengal and East India, the simulated geopotential height at 600 hPa of four ensemble members is shown in Figure S6. %\ref{fig:FIG15}.
Apart from a slight deepening of height around 96-120 hours of model integration (Figure S6), none of the ensemble members shows a strong signature of either LPS in Bay of Bengal or an MTC over Arabian sea. This suggest that in the absence of Bay of Bengal LPS Arabian sea system does not form.  Further for robustness the composite difference of geopotential height of control and cold-dry Bay of Bengal is shown in Figure \ref{fig:FIG13}. This clearly suggests that favorable conditions over the Bay of Bengal allow the intensification of Bay of Bengal LPSs, which in turn deepens the trough over the Arabian Sea and supports the MTC formation over western India and the northeast Arabian Sea. This indicates that with the unfavorable conditions over the Bay of Bengal, the LPS did not intensify, and consequently, the MTC did not form over the Arabian Sea either. %These experiments, supported by observations of Type 2a composites, suggest that the Bay of Bengal system enhances the favorable conditions in forming Type 2a Arabian Sea MTCs. 
In more detail, Figure S7 displays the time series of absolute vorticity at 600 hPa, specifically averaged over the Arabian Sea MTCs, for each ensemble member in both the control and cold-dry Bay of Bengal simulations. Notably, the vorticity in the northeast Arabian Sea consistently remained lower than that in the control run throughout the entire simulation period. Furthermore, the cold-dry Bay of Bengal experiments show a continuous decrease in absolute vorticity. This indicates unfavorable conditions over the Arabian Sea compared to the control run, despite only forcing the conditions in the Bay of Bengal unfavorable. These findings strongly suggest that the presence of LPS in the Bay of Bengal, along with its remote effects, plays a crucial role in regulating the distribution of vorticity, horizontal shear and middle troposphere moisture over the Arabian Sea. Consequently, the frequent coexistence of LPSs in the Bay of Bengal and MTCs over the Arabian Sea cannot be regarded as a mere coincidence. Instead, it implies that the presence of Bay of Bengal LPSs, its associate convective heating and remote response play a critical role in creating favorable conditions for the formation and, to some extent, the maintenance of Arabian Sea MTCs.

\noindent However, it should be kept in mind that although the majority of in-situ Type-2a MTCs are influenced by the Bay of Bengal system, not all MTCs in the Arabian Sea necessarily require the involvement of the Bay of Bengal system. In fact, Type 2b MTCs (Kushwaha et al., 2023), which form during the early monsoon season, are generated by the monsoon vortex, while Type 2c MTCs (Kushwaha et al., 2023) also undergo in-situ genesis but originate from precursors in the southern Bay of Bengal. It is worth noting that Type 2c MTCs are the weakest category among all MTCs and form under unfavorable conditions over the Bay of Bengal \cite{kushwaha2023classification}. This observation further emphasizes the significance of the Bay of Bengal convection and associated remote teleconnection in the majority of rainy {\it in-situ} MTCs in the Arabian Sea.
%===================================================

\section{Conclusions}
A consistent theme observed in previous research, starting from the seminal work of \citet{miller1968iioe} to more recent studies by \citet{choudhury2018phenomenological} and the comprehensive objective classification Mid-Troposphere Cyclones (MTCs) by \citet{kushwaha2023classification}, is the close association between rainy MTCs over western India and the presence of low-pressure systems (LPS) over the Bay of Bengal. It has been observed that, in many instances, the LPS in the Bay of Bengal precedes the formation of MTCs in the Arabian Sea. Through a systematic set of numerical experiments presented in this study, we have demonstrated that this coexistence is not coincidental; rather, the Bay of Bengal systems plays a crucial role in the genesis of Arabian Sea MTCs.

\noindent We initiated the analysis by examining composite of MTC genesis events where LPS in the Bay of Bengal preceded and coexisted with the Arabian sea MTCs (referred to as Type 2a MTCs in \citet{kushwaha2023classification}). Conducting a lag composite of dynamical and thermodynamic fields for Type 2a MTCs revealed that the formation and intensification of Bay of Bengal systems accompanied a westward-extending middle troposphere trough. This trough, akin to an off-equatorial heat-induced Gill-type response, extended from the Bay of Bengal to the Arabian Sea.
The middle troposphere zonal trough enhanced the horizontal zonal shear over western India and established middle tropospheric easterlies over northwest India. These middle troposphere easterlies prevented the mixing of dry and hot air from the desert regions to the north and west, resulting in a depletion of the low-level inversion layer and destabilization of the lower troposphere. Consequently, the moistening of the middle troposphere occurred over the Arabian Sea and western India. Given that this region is typically unfavorable for moist convective activity climatologically, the moistening of the middle troposphere creates favorable conditions for deep convection to occur.
These alterations in the dynamic and thermodynamic environment over the Arabian Sea, triggered by the presence of Bay of Bengal systems, provide a fertile ground for the genesis and growth of mid-troposphere cyclones. 

\noindent The apparent link between Arabian Sea MTC formation and Bay of Bengal LPSs was further explored with numerical experiments. Two sets of experiments were performed: first, with added "Bogus Vortices or LPS" over the Bay of Bengal on top of June-July climatology, and second, the effect of preventing the formation or weakening of Bay of Bengal LPSs on MTC formation was studied by cooling and drying the Bay of Bengal. Interestingly, in most of the ensemble members when bogus vortices are introduced over the specific locations of Bay of Bengal (where LPSs are most frequently seen before Type 2a MTC formation in observation), the genesis of MTC takes place over Arabian Sea in 2.5--4 days of model simulations within the induced middle troposphere trough.  With the addition of a Bogus Vortex over the Bay of Bengal, a westward extending trough forms stretching from the Bay of Bengal up to the Arabian Sea; as a result, easterlies are established north of the trough axis (around $20^{\circ}$N) and westerlies to the south. This modified flow enhances the horizontal shear and background vorticity. Middle troposphere easterlies north of the trough prevent dry air intrusion from the west and north, reduce the low-level inversion, and destabilize the lower troposphere. These features in the numerical simulations agree with the reanalysis-based characterization of the role of Bay of Bengal LPS in the development of Type 2a MTCs.  Furthermore, the vorticity budget of induced MTCs shows that during the first 24 hours of MTC formation, advection of absolute vorticity and tilting account for the intensification. However, during the rapid intensification phase, vortex stretching dominates. The $\beta$-term remains positive throughout; however, it contributes mainly in the early phase of intensification. Vertical advection is initially quite weak, but it does contribute to the vorticity tendency in the middle troposphere in the later stages of MTC development.

\noindent A mechanism denial experiment involving the cooling and drying of the Bay of Bengal was conducted to confirm the link between Low-Pressure Systems (LPS) in the Bay of Bengal and the genesis of Mid-Troposphere Cyclones (MTCs) in the Arabian Sea. The experiment consisted of a control run where an actual Arabian Sea MTC was successfully simulated following the formation of an LPS over the Bay of Bengal. Subsequently, the Bay of Bengal was artificially cooled and dried to suppress and weaken the formation of Bay of Bengal LPSs.

In the control experiment, the LPS over the Bay of Bengal intensified, leading to the development of a westward-extending trough. This resulted in the strengthening of horizontal shear and the establishment of easterly winds. Consequently, an MTC formed over the Arabian Sea as expected. However, in the experiment with the cooled and dried Bay of Bengal, LPS over the Bay of Bengal failed to intensify. As a result, the westward-extending trough did not develop, and the horizontal shear and easterly winds remained weak. Consequently, an MTC did not form over the Arabian Sea. These findings provide compelling evidence supporting the notion that the coexistence of the Bay of Bengal system during the formation of Arabian Sea MTCs is not a mere coincidence. Instead, the genesis and maintenance of the majority of Type 2a MTCs can be attributed directly to the dynamic influence of LPS over the Bay of Bengal. The mechanism denial experiment, by dampening the formation of Bay of Bengal LPSs through cooling and drying, clearly demonstrated the crucial role of these systems in the development of MTCs over the Arabian Sea.
%It is important to not that though the gensis of Type-2c which is also a in-situ formation category and does not require any active Bay of Bengal systems however is precursors originates near Sri-Lanka regions unlike Type-2a which forms locally over Arabian sea in the presence of Bay of Bengal system without any link to remote precursors. Thought the Type-2a, Type-2b and Type-2c all are the category of in-situ gensis, Type-2a directly affected  by the presence of Bay of Bengal systems in view of absence of its remote precursors. }

%\clearpage

% \begin{center}
% {\Large {\bf Supplementary Material}}
% \end{center}
\supplementarysection
%=================================================
\newpage
\begin{figure}
\centering
\includegraphics[trim=0 0 0 0, clip, height = 1\textwidth, width = 1\textwidth, angle = 90, clip]{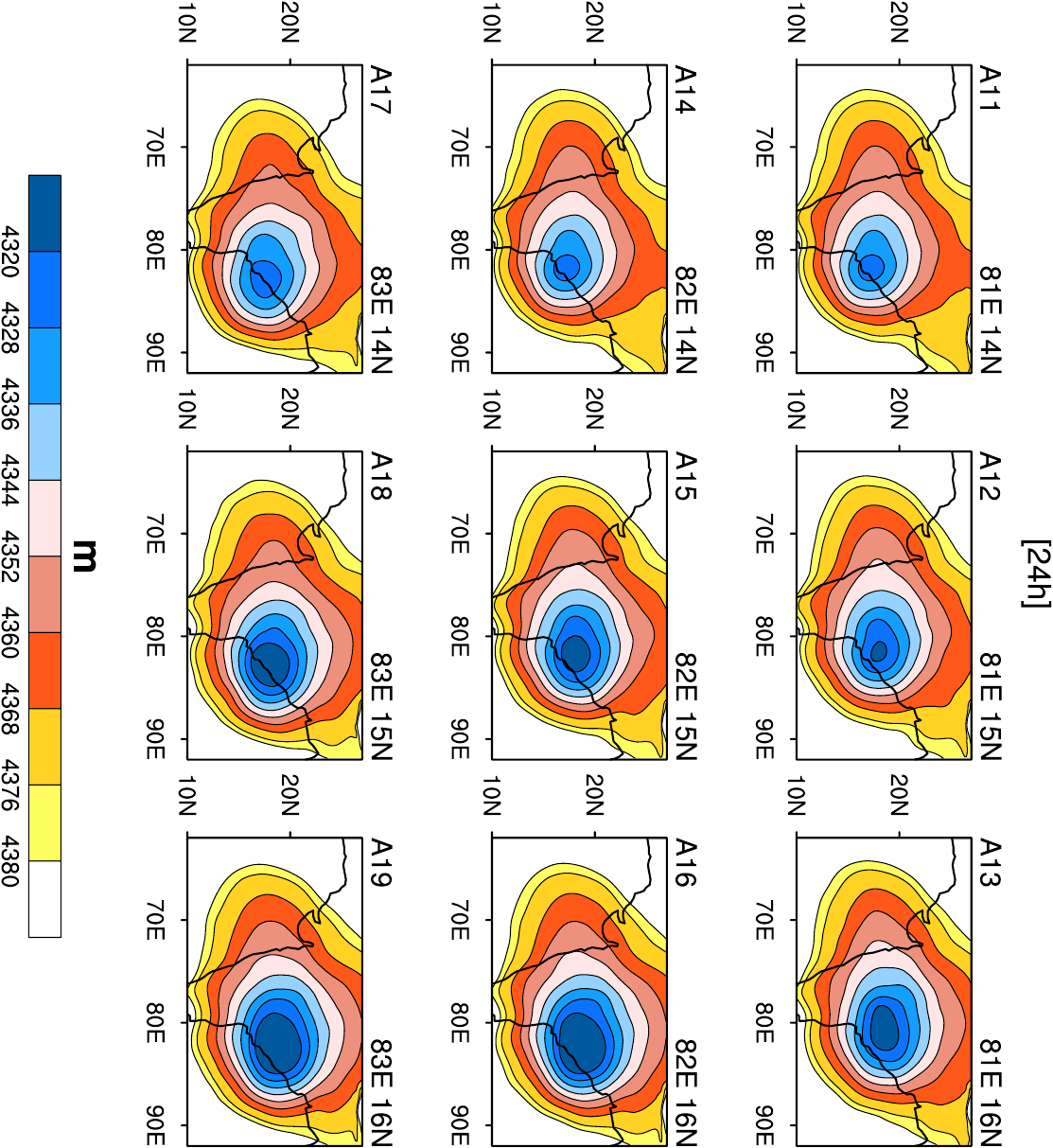}
\caption{Geopotential height at 600 hPa of nine members (A11-A19) of group A1 at 24 hours of simulation.}
%\setfigurenum{S1} 
\label{fig:S1}
\end{figure}
%=================================================

%=================================================
\begin{figure}
\centering
\includegraphics[trim=0 0 0 0, clip, height = 1\textwidth, width = 1\textwidth, angle = 90, clip]{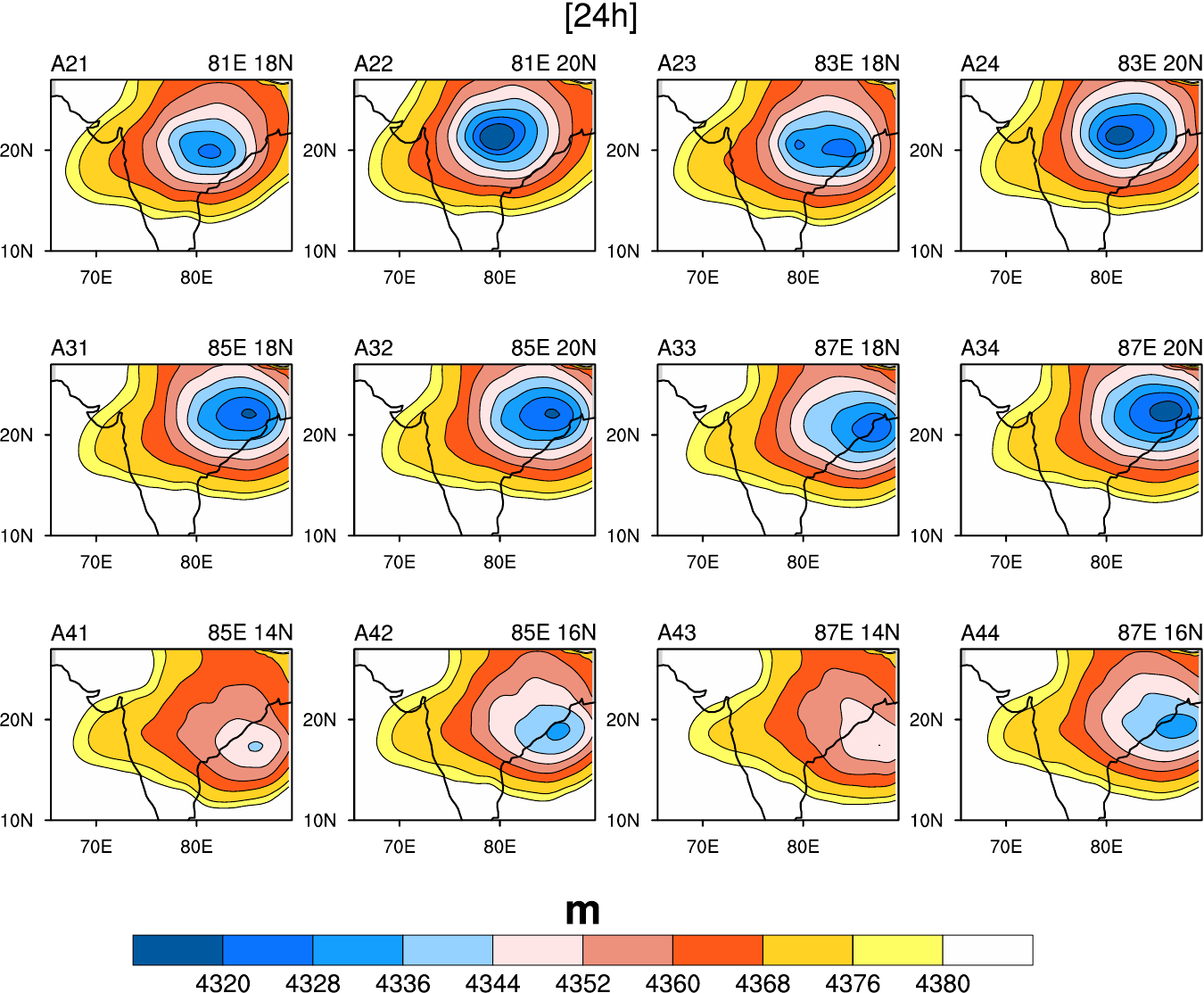}
\caption{Geopotential height at 600 hPa of nine members (A11-A19) of group A1 at 24 hours of simulation.}
%\setfigurenum{S2} 
\label{fig:S2}
\end{figure}
%==================================
\begin{figure}   
\vspace{-0.3cm}
\centering
\includegraphics[trim=0 0 0 0, clip, height = 0.8\textwidth, width = 1\textwidth, angle = 0, clip]{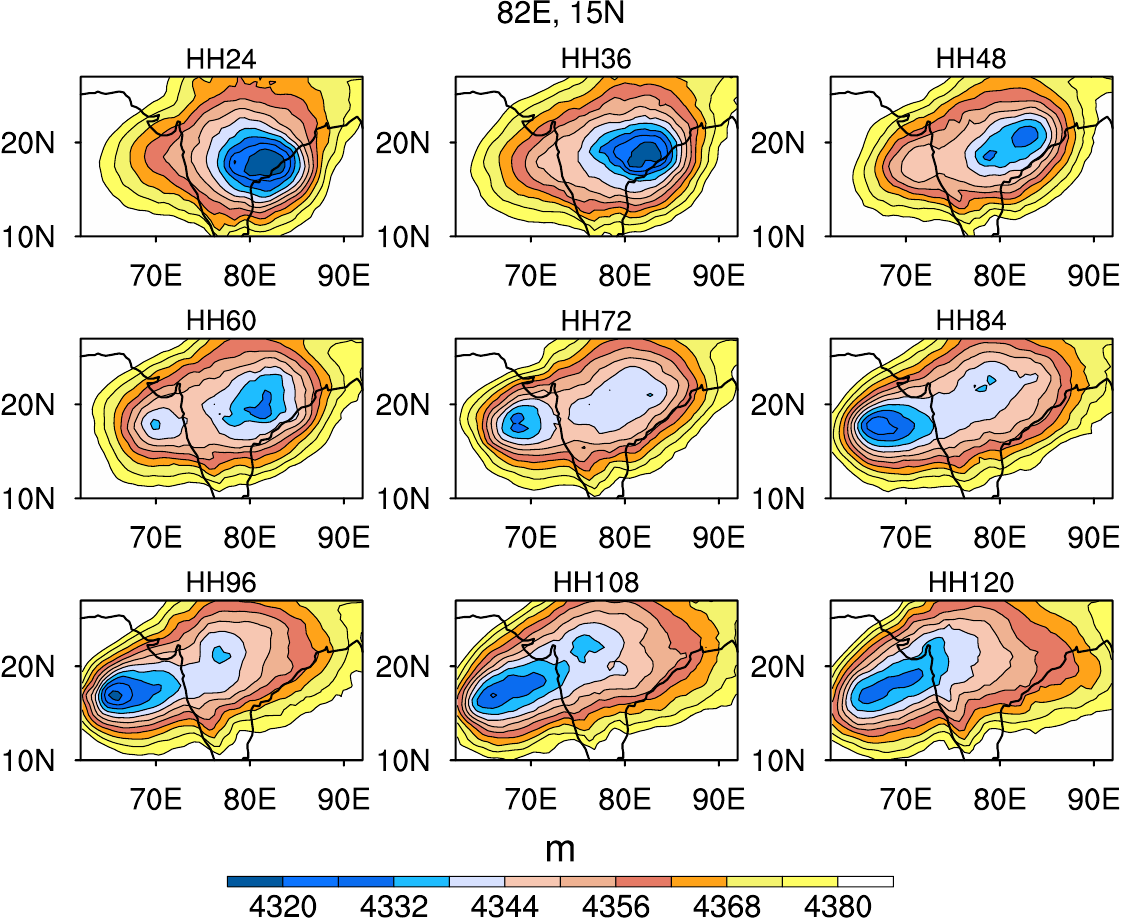}

\caption{The figure presents the geopotential height at the 600 hPa level for an ensemble member initialized with the addition of a bogus vortex at 82E and 15N. The simulations cover a time span from 24 to 120 hours from  left to right. }
%\setfigurenum{S3}
\label{fig:S3}
\end{figure}
%\selectlanguage{english}
%\bibliography{bibliography/converted_to_latex.bib%
%}

%============================

%%%%%%%%%%%%%%%%%%%%%%%%%%%
\begin{figure}   
\vspace{-0.3cm}
\centering
\includegraphics[trim=0 0 0 0, clip, height = 0.4\textwidth, width = 0.8\textwidth, angle = 0, clip]{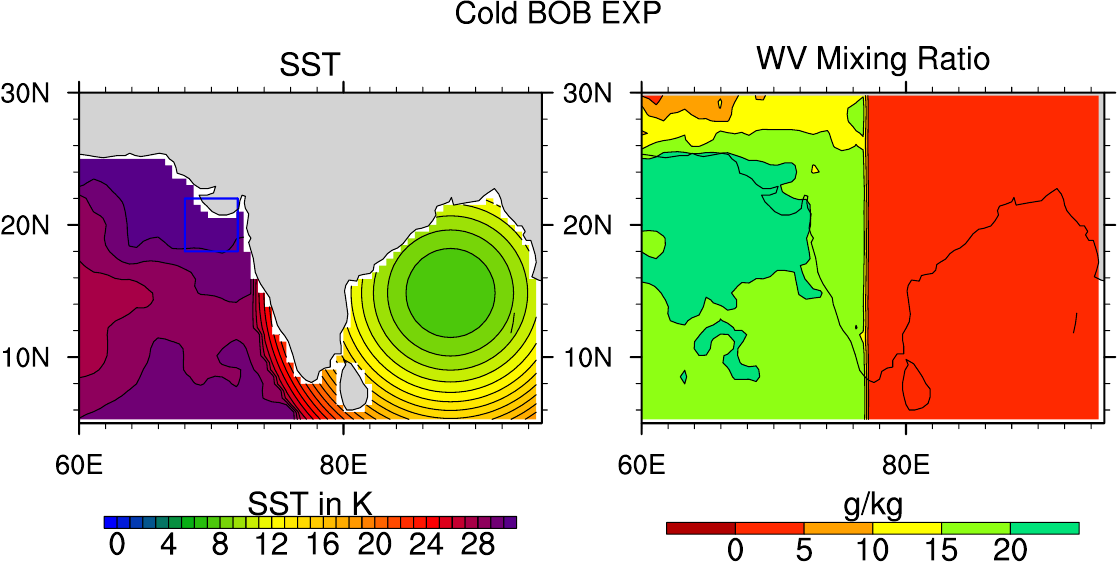}
\caption{Initial conditions  (a) cold Gaussian SST blob over the Bay of Bengal, (b) Water Vapor (WV) mixing ratio at 1000 hPa --- dry east India and BOB. Blue box in Figure-a represents the area used to calculate the average in Figure-S8}
\label{fig:S4}
%\setfigurenum{S4}
\end{figure}
%%%%%%%%%%%%%%%%%%%%%%%%%%%%%%

%===================================================
\begin{figure}   
\vspace{-0.3cm}
\centering
\includegraphics[trim=0 0 0 0, clip, height = 0.8\textwidth, width = 1\textwidth, angle = 0, clip]{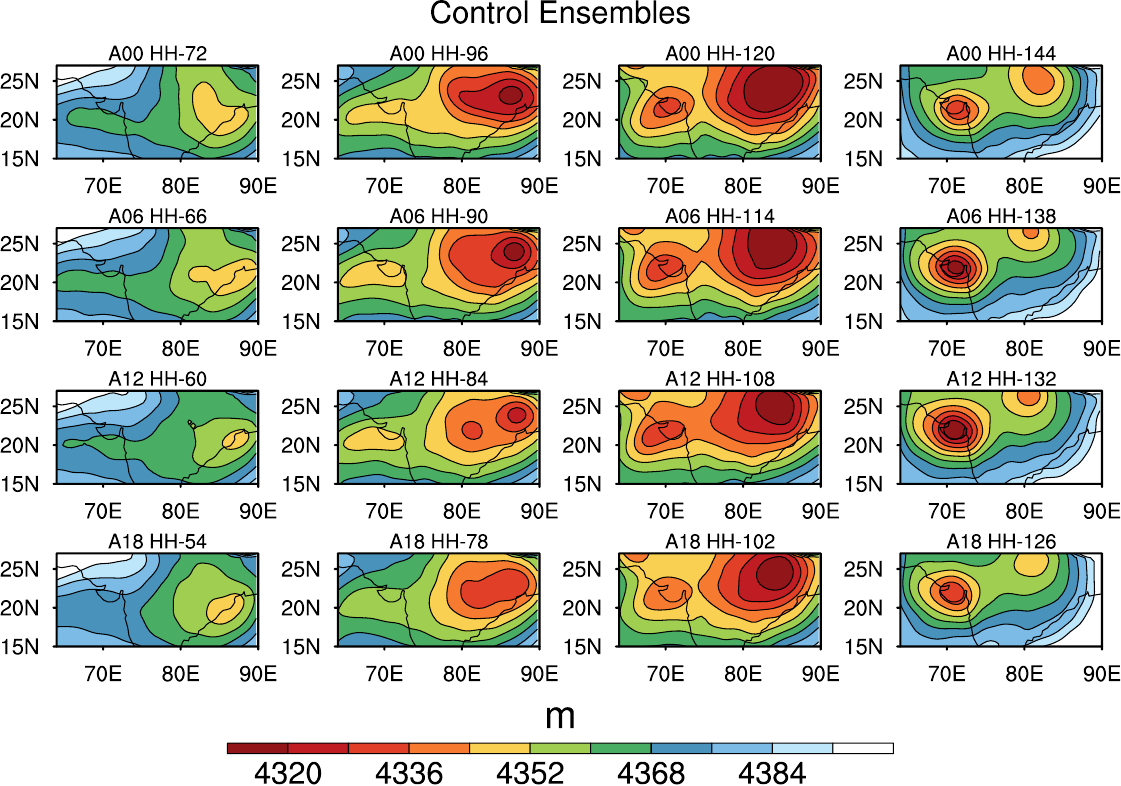}

\caption{Geopotential height at 600 hPa of the control ensemble members for the July 2020 MTC. Rows 1 to 4 represent simulations initialized on 1 July at 00, 06, 12, and 18 UTC, respectively. From left to right, each sub-panel shows snapshots as the simulation time progresses. In row 1, the snapshots range from 72 to 144 hours, in row 2 from 66 to 130 hours, in row 3 from 60 to 132 hours, and in row 4 from 54 to 126 hours. }
\label{fig:S5}
%\setfigurenum{S5}
\end{figure}
%%%%%%%%%%%%%%%%%%%%%%%%%%%

%%%%%%%%%%%%%%%%%%%%%%%%%%%
\begin{figure}   
\vspace{-0.3cm}
\centering
\includegraphics[trim=0 0 0 0, clip, height = 0.8\textwidth, width = 1\textwidth, angle = 0, clip]{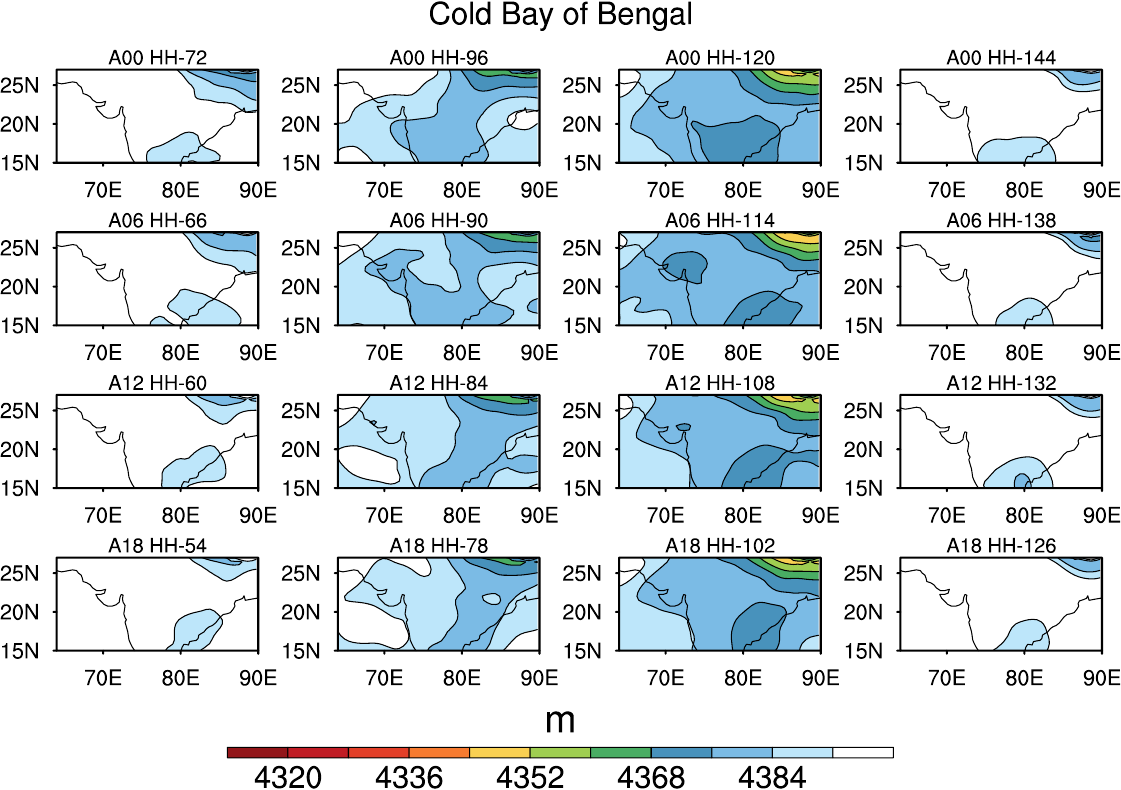}
\caption{Geopotential height at 600 hPa of the cold-dry BOB ensemble members for the July 2020 MTC. Rows 1 to 4 represent simulations initialized on 1 July at 00, 06, 12, and 18 UTC, respectively. From left to right, each sub-panel shows snapshots as the simulation time progresses. In row 1, the snapshots range from 72 to 144 hours, in row 2 from 66 to 138 hours, in row 3 from 60 to 132 hours, and in row 4 from 54 to 126 hours.}
%\setfigurenum{S6}
\label{fig:S6}
\end{figure}

% \begin{figure}   
% \vspace{-0.3cm}
% \centering
% \includegraphics[trim=0 0 0 0, clip, height = 0.8\textwidth, width = 1\textwidth, angle = 0, clip]{CONTROL_SIMULATIONS-crop.pdf}

% \caption{The figure displays the geopotential height at the 600 hPa level for a simulation utilizing climatological initial conditions without the inclusion of any additional vortex. The simulations span a duration of 24 to 120 hours. The six-hourly climatology was calculated based on data from 20 June to 10 July, covering the years 2000 to 2015.}
% \setfigurenum{S7}
% \label{fig:FIG15}
% \end{figure}

\begin{figure}   
\vspace{-0.3cm}
\centering
\includegraphics[trim=0 0 0 0, clip, height = 0.9\textwidth, width = 0.8\textwidth, angle = 90, clip]{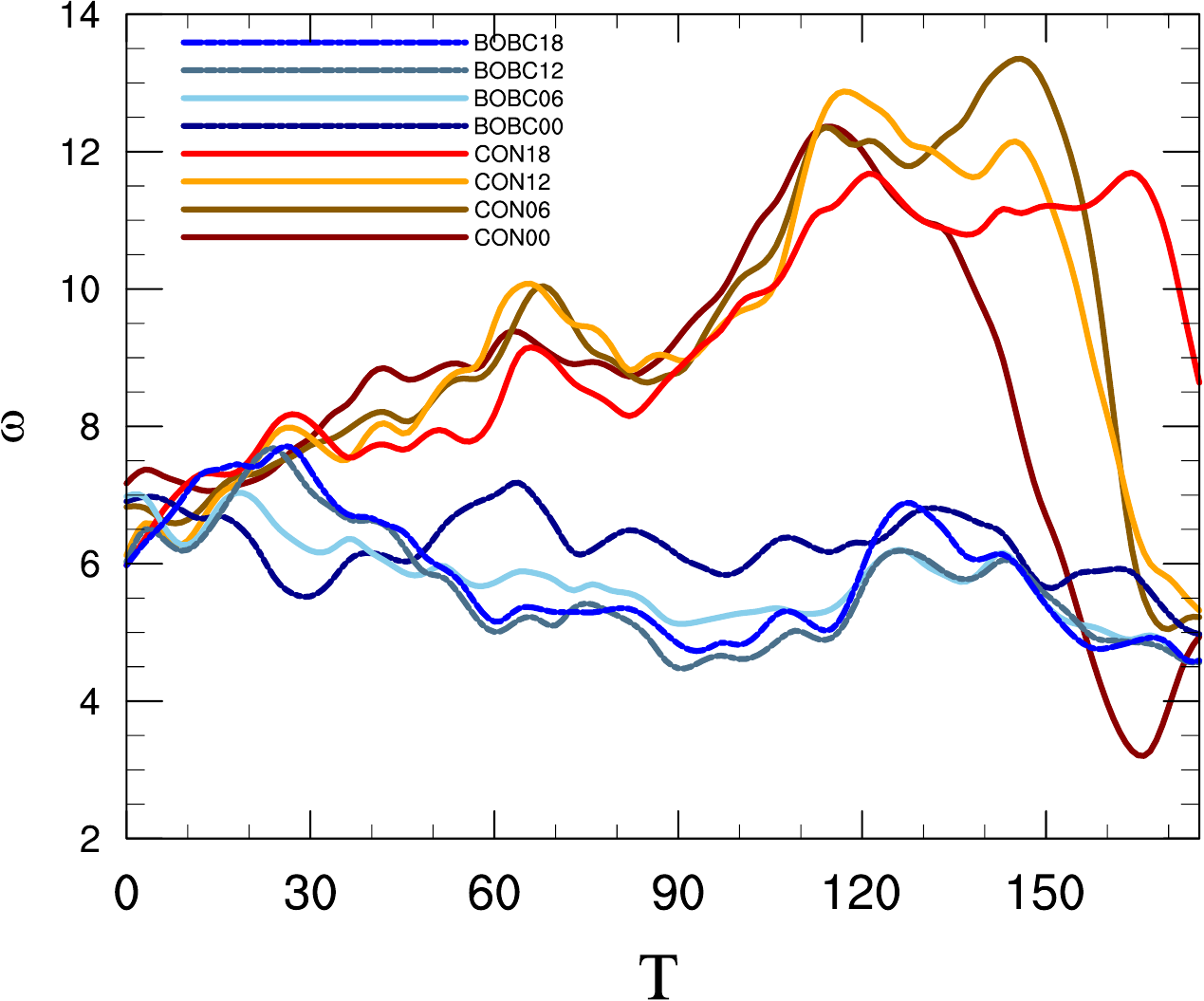}
\caption{Time series of 600 hPa averaged absolute vorticity ($10^{-5}s^{-1}$) in the northeastern Arabian Sea (18$^{\circ}$N--23$^{\circ}$N and 68$^{\circ}$E--72$^{\circ}$E), within the indicated blue box (Figure S4). It showcases the control (CON) and cold-dry Bay of Bengal (BOBC) experiment for the July 2020 case. The group of coloured lines represents the initialization conducted at 00, 06, 12, and 18 UTC, respectively or both control and dry-cold BOB case. The abscissa represents the simulation time in hours, starting from 1 July 18 UTC (T=0 in figure) for all members.}
%\setfigurenum{S7}
\label{fig:S7}
\end{figure}

%\section{References}
\clearpage 

\bibliographystyle{apalike}
\bibliography{ref.bib}

\end{document}